\documentclass[11pt,a4paper]{article} 
\usepackage{jheppub}

\usepackage{slashed}
\usepackage{youngtab}
\usepackage{amsmath}

\usepackage{float}

\newcommand{\beq}{\begin{equation}}
\newcommand{\eeq}{\end{equation}}

\newcommand{\beqa}{\begin{eqnarray}}
\newcommand{\eeqa}{\end{eqnarray}}

\newcommand{\by}{\begin{eqnarray}}
\newcommand{\ey}{\end{eqnarray}}
\newcommand{\al}{\alpha}
\newcommand{\be}{\beta}
\newcommand{\ga}{\gamma}

\newcommand{\de}{\delta}
\newcommand{\ep}{\epsilon}

\newcommand{\si}{\sigma}
\newcommand{\half}{\frac{1}{2}}

\newcommand{\thalf}{\tfrac{1}{2}}

\newcommand{\mN}{\mathcal N}

\newcommand\fverb{\setbox\fverbbox=\hbox\bgroup\verb}
\newcommand\fverbdo{\egroup\medskip\noindent%
            \fbox{\unhbox\fverbbox}\ }
\newcommand\fverbit{\egroup\item[\fbox{\unhbox\fverbbox}]}
\newbox\fverbbox

\newcommand{\nablaslash}{\not{\hbox{\kern-3pt $\nabla$}}}

\title{On the squashed seven-sphere operator spectrum}

\author[a]{Simon Ekhammar}
 \author[b]{and Bengt E.W.~Nilsson}
 \affiliation[a]{Department of Physics and Astronomy\\
Uppsala University, Box 516\\
SE-751 20 Uppsala, Sweden}
\affiliation[b]{Department of Physics \\ Chalmers University of Technology\\
SE-412 96 G\"oteborg, Sweden} 
 \emailAdd{simon.ekhammar@physics.uu.se, tfebn@chalmers.se}

\abstract{ We derive major parts of  the eigenvalue spectrum of the operators on the squashed  seven-sphere that appear in the compactification of eleven-dimensional supergravity.
These spectra determine the mass spectrum of the fields in $AdS_4$ and are important for the corresponding  $\mN =1$ supermultiplet structure. 
This work is a continuation of the work in \cite{Nilsson:2018lof} where the complete spectrum of irreducible isometry  representations of the fields in $AdS_4$ was derived 
for this compactification. Some comments are also made concerning  the $G_2$  holonomy
 and its implications  on the structure of the operator equations on the squashed  seven-sphere.}


\toccontinuoustrue
\begin{document}
\maketitle


\section{Introduction}

The purpose of this paper is to continue the work in \cite{Nilsson:2018lof} and derive the eigenvalue spectrum of some of the operators that determine the masses and supermultiplet structure
of the entire Kaluza-Klein spectrum of the squashed seven-sphere compactification of eleven-dimensional supergravity. 
The list of previously known eigenvalue spectra, containing $\Delta_0$, ${\slashed D}_{1/2}$ and $\Delta_1$, is in this paper extended by those of
$\Delta_2$, $\Delta_3$ and $\Delta_L$ while  ${\slashed D}_{3/2}$ remains to be done.
Note that parts of the spectrum of the Lichnerowicz operator $\Delta_L$ are derived below but they were quoted already in \cite{Duff:1986hr}\footnote{See Ref. [198], unpublished work by Nilsson and Pope.}.

With the intention to keep this paper as  brief as possible we refer the reader to the review \cite{Duff:1986hr} for a Kaluza-Klein background on the problem and 
for some of the necessary details and conventions needed in the derivations below. For the full structure of irreducible isometry representations on $AdS_4$ stemming from
 the squashed seven-sphere compactification we
refer to \cite{Nilsson:2018lof}. The latter paper summarises in a few pages  the most relevant information from  \cite{Duff:1986hr} in particular several tables
that are spread out in various chapters of  \cite{Duff:1986hr}.
In this spirit we present below just the most crucial formulas that are needed to define the problem and derive the spectra. In the Appendix we collect a number of useful octonionic identities and other formulas, 
as well as some for our purposes crucial Weyl tensor calculations.

The coset description  of the squashed seven-sphere is 
\beq
G/H=Sp_2\times Sp_1^C/Sp_1^A\times Sp_1^{B+C},
\eeq
where, in order to define the denominator, $Sp_2$ is split into $Sp_1^A\times Sp_1^B$ and $Sp_1^{B+C}$ is the diagonal subgroup of $Sp_1^B$ and $ Sp_1^C$. The $G$ irreps specifying the mode functions $Y$
of a general Fourier expansion on $G/H$ are denoted $(p,q;r)$ \cite{Duff:1986hr} and the entire Kaluza-Klein irrep spectrum is derived and tabulated in terms of {\it cross diagrams}  in \cite{ Nilsson:2018lof}. 
The present paper will fill in some gaps in our knowledge of the eigenvalue spectrum of the relevant operators on the squashed seven-sphere but some are still missing. The remaining issues that we need to  address to complete the eigenvalue spectra will be elaborated upon in a forthcoming publication \cite{Nilsson:2021kn}.

The interest in deriving complete spectra in various Kaluza-Klein compactifications has recently been revived due to the discovery of new powerful versions of the embedding tensor technique.
However, these methods can be applied directly only when the vacuum is an extremum of a maximal gauged supergravity theory in $AdS_4$ which can be lifted to ten or eleven dimensions,
see for instance \cite{Malek:2019eaz, Cesaro:2021haf} and references therein\footnote{We are grateful to Oscar Varela for a clarification on this issue.}. The point we want to emphasise here is  that, due to the {\it space invaders scenario} \cite{Duff:1983ajq, Duff:1986hr}, the squashed seven-sphere solution is not of this kind and it therefore 
seems clear that other methods
are required to obtain the $AdS_4$ spectrum in this case.

There are also potential applications of this work in the context of the swampland program\footnote{We are grateful to M.J. Duff for raising some interesting aspects of this question that eventually led to this work.}, 
see for instance \cite{Vafa:2005ui, ArkaniHamed:2006dz, Ooguri:2016pdq} 
and references therein. This particular connection will be addressed elsewhere.

In the next section we very briefly review the background of the problem and the  method that is used in this paper to derive the eigenvalue spectrum of 
operators on coset manifolds. The method is explained in many places, e.g. \cite{Salam:1981xd, Duff:1986hr},
 and in Section 3 we first apply it to obtain the spectra of the spin 0 and 1 operators giving 
straightforwardly the well known results cited in \cite{Duff:1986hr}.
 Already when applied to the spin 1/2 Dirac operator complications arise
which get further pronounced when we subsequently turn to more and more complicated operators.
A summary of the obtained eigenvalues is provided in the Conclusions, together with comments on some of the remaining issues.
Some technical aspects needed in the derivations below are explained  in the Appendix.

%
%
\section{Compactification on the seven-sphere}

The Fourier expansion technique that we will apply is, following the general strategy explained in \cite{Salam:1981xd} and summarised in \cite{Duff:1986hr}, 
based on the  {\it coset master equation} for the $Spin(n)$ covariant derivative
on a $n$-dimensional coset space $G/H$:
\beq
\label{cma}
\nabla_a Y+\thalf f^{bc}{}_a\,\Sigma_{bc} Y=-T_a Y.
\eeq
The mode functions $Y$ have two suppressed indices: 
One corresponding to the $Spin(n)$ tangent space irrep of the field that is being Fourier expanded\footnote{Note, however, that the whole  matrix of the $G$ group element is involved in this equation.} and one that specifies the mode, that is a $G$ irrep. This  equation 
is derived in \cite{Duff:1986hr}  from group elements of the isometry group $G$. In the conventions used there the Lie algebra of the group $G$ has generators $T_A$, satisfying $[T_A, T_B]=f_{AB}{}^CT_C $, 
which are divided
into $T_{\bar a}$ of the subgroup $H$ and $T_{a}$ in the complement of $H$ in $G$. Thus, if the Lie algebra of $G$ is reductive it  splits as follows:
\beq
\label{reductiveliealg}
[T_{\bar a}, T_{\bar b}]=f_{\bar a \bar b}{}^{\bar c}T_{\bar c},\,\,\,[T_{\bar a}, T_{b}]=f_{\bar a b}{}^{c}T_{c},\,\,\,\,\,\,\,[T_{a}, T_{b}]=f_{a b}{}^{\bar c}T_{\bar c}+f_{a b}{}^{c}T_{c}.
\eeq
The indices $a,b,..$ are vector indices in the tangent space of $G/H$ and $\nabla_a$ in (\ref{cma}) is an $Spin(n)$  covariant derivative
with a torsion free spin connection while the $\Sigma_{ab}$ are the generators of $Spin(n)$ in the representation relevant for the operator equation we are solving.
Note that it is only the structure constants $f_{a b}{}^{c}$ that appear in the coset master equation (\ref{cma}). For symmetric spaces like the round seven-sphere $f_{a b}{}^{c}=0$.
Furthermore, the second algebraic relation in (\ref{reductiveliealg}) defines the imbedding of $H$ in the tangent space group $Spin(n)$.
 
For symmetric coset spaces the coset master equations thus reads $\nabla_a Y=-T_a Y$
and the eigenvalue spectrum for the operators on $G/H$ are rather easily derived. The complications arising in the squashed seven-sphere case therefore stem
entirely from the structure constant dependent term in (\ref{cma}). As first shown in \cite{Bais:1983wc} for the squashed seven-sphere they are  given by 
 the structure constants of the octonions $a_{abc}$. In the normalisation used in \cite{Duff:1986hr} they read
\beq
f_{abc}=-\tfrac{1}{\sqrt{5}}a_{abc}=-\tfrac{2}{3}m\,a_{abc}.
\eeq
Since octonions will play a key role in the rest of this paper we have listed some useful octonionic identities  in the Appendix. Furthermore, in the last expression
 for these structure constants  we have introduced the scale parameter $m$ arising from the ansatz $F_{\mu\nu\rho\si}=3m\ep_{\mu\nu\rho\si}$ in the compactification of eleven-dimensional supergravity. 
 This is useful since we are dealing with  dimensionful quantities, like the covariant derivatives and Hodge-de Rham operators. The conventions used in \cite{Duff:1986hr} correspond to setting
\beq
m^2=\tfrac{9}{20}.
\eeq  
We will assume $m>0$ and also use $m=\tfrac{3}{2\sqrt{5}}$ as for $f_{abc}$ above. 
The squashed seven-sphere discussed in this work has the orientation that gives rise to $\mN =1$ supersymmetry in $AdS_4$ after compactification. This fact will be used below
when we discuss the corresponding supermultiplets and the $SO(3,2)$ irreps entering these multiplets.

With this preparation we can readily attack the eigenvalue problems $\Delta_p Y_p = \kappa_p^2 Y_p$ where the Hodge-de Rham\footnote{This operator generalises the  Laplace-Beltrami operator to $p$-forms with $p>0$.} 
operator on $p$-forms is defined as $\Delta_p = \text{d}\delta + \delta \text{d}$. Here $\text{d}$ is the exterior derivative and $\delta=(-1)^p\star \text{d} \star$ its adjoint. These operators act on forms according to
\beq
	(\text{d}Y)_{a_1\dots a_p} = p\nabla_{[a_1}Y_{a_2\dots a_p]},
	\quad
	(\delta Y)_{a_1\dots a_p} = -\nabla^{b}Y_{ba_1\dots a_p},
\eeq
where brackets here and in the following are weighted antisymmetrisations, $Y_{[a_1\dots a_p]} = \frac{1}{p!}(Y_{a_1\dots a_p}+\text{permutations with sign})$. The explicit form of $\Delta_p$ 
can for all operators considered in this paper be expressed using the Riemann tensor, $R_{abcd}$, and the d'Alembertian $\Box \equiv \nabla^a\nabla_a$. 
On the squashed seven-sphere, which is an Einstein space with $R=42m^2$, 
\beq\label{eq:RiemannDef}
R_{ab}{}^{cd}=W_{ab}{}^{cd}+2m^2\de_{ab}^{cd},
\eeq 
where $W_{abcd}$ is the Weyl tensor given in the Appendix. 

The main use of the coset master equation \eqref{cma} will be to replace derivatives by group-theoretical algebraic data. In particular,  for the squashed sphere, by  squaring \eqref{cma} 
 we obtain an expression for $\Box$ which can then be used to replace $\Box$ by algebraic data plus terms linear in $\nabla_a$. This will be clear below.

\section{Eigenvalue spectra on the squashed seven-sphere}
In this section we start by  briefly reviewing some of the results previously obtained with the coset space techniques mentioned above.
These results are then generalised to the more complicated operators for which we present   a number of  new eigenvalues. 

\subsection{Review of the method applied to forms of rank 0 and 1}

The method outlined above becomes rather trivial when applied to 0-forms. In this case we want to solve the eigenvalue equation
\beq
\Delta_0 Y=\kappa_0^2 Y,
\eeq
where the positive semi-definite Hodge-de Rham operator  on 0-forms is $\Delta_0=-\Box$. Thus, following \cite{Duff:1986hr},  the spectrum is obtained directly by squaring the coset master equation (\ref{cma}):
\beq
\Delta_0 Y=-\Box Y=-T_aT_aY=(C_G-C_H)Y=\kappa_0^2 Y,
\eeq
where the two second order Casimir operators for $G=Sp_2\times Sp_1^C$ and $H=Sp_1^A\times Sp_1^{B+C}$  have eigenvalues 
\beq
C_G=C(p,q)+3C^C(r)=\thalf(p^2+2q^2+4q+6q+2pq)+\tfrac{3}{4}r(r+2),
\eeq
and
\beq
C_H=2C^A(s)+\tfrac{6}{5}C^{B+C}(t)=\thalf\,s(s+2)+\tfrac{3}{10}\,t(t+2).
\eeq
Here $(p,q)$ are Dynkin labels for $Sp_2$ irreps and $r,s,t$ those for the   irreps of the various $Sp_1$ groups that occur in the squashed seven-sphere  coset description. 

The whole problem of deriving the eigenvalue spectrum of  harmonics is reduced to determining first which $H$ irreps are contained
in the $Spin(7)$ irrep of Y (trivial for 0-forms) and then finding all $G$ irreps that when split into $H$ irreps contain any of the $H$ irreps found in the $Spin(7)$ irrep of Y. 
Some partial  results on the spectrum of $G$ irreps are obtained in \cite{Nilsson:1983ru, Duff:1986hr} by breaking up the spectrum on the round sphere according to $Spin(8)\rightarrow Sp_2\times Sp_1$ while the entire spectrum
of all operators is derived directly from the squashed sphere coset in \cite{Nilsson:2018lof} using the coset method just described.

For  0-forms $C_H=0$  and the spectrum therefore becomes $\kappa^2_0=C_G$. 
However, as explained above,  we should introduce the scale parameter $m$.  In view of this we can use  $m^2=\tfrac{9}{20}$ to write the
 eigenvalues  as follows:\footnote{We will in the rest of this paper display eigenvalues either as $\kappa_p^2$ or, equivalently, as $\Delta_p$.}
\beq
\Delta_0=\kappa^2_0=\frac{m^2}{9} 20\,C_G(p,q;r).
\eeq
As an example how this result is used let us consider the spectrum in the graviton sector\footnote{As in \cite{Nilsson:2018lof} we will use the notation of  \cite{Breitenlohner:1982jf} which includes the parity  of the field, as in, e.g.,  $2^+$.}: In units of $m$  the $SO(3,2)$ irrep defining energy $E_0(spin)$ is \cite{Heidenreich:1982rz, Breitenlohner:1982bm, Englert:1983rn}
\beq
E_0(2^+)=\frac{3}{2}+\half\sqrt{\frac{M_2^2}{m^2}+9}=\frac{3}{2}+\half\sqrt{\frac{20C_G}{9}+9}=\frac{3}{2}+\frac{1}{6}\sqrt{20C_G+81},
\eeq
where we have used the fact that the mass-square operator for spin 2 fields in $AdS_4$ is $M^2_2=\Delta_0$.
We will occasionally refer to  results like this as being of {\it square-root  form}, here a $\sqrt{81}$-form. The reason for this is that all the fields in a supermultiplet must be of the same square-root  form.

We now turn to the 1-form case and review the result and the derivation in \cite{Duff:1986hr}\footnote{For a different derivation see \cite{Yamagishi:1983ri}.}. In this case the structure constant term in the coset master equation comes in and complicates the 
calculation somewhat. As for the 0-form modes we write out the $\Delta_1$ eigenvalue equation explicitly:
\beq
\Delta_1:\,\,\,\,\Delta_1 Y_a = - \Box Y_a + R_a{}^bY_b=-\Box Y_a+ 6m^2 Y_a = \kappa_1^2 Y_a,
\eeq
We want to use the square of the  coset master equation to eliminate the $\Box$ term: Inserting $f_{abc}=-\tfrac{2}{3}ma_{abc}$
 into \eqref{cma} gives $\nabla_aY_b-\frac{m}{3}a_{abc}Y_c=-T_aY_b$ which when squared yields
\beq
G/H:\,\,\,\,\,\Box Y_a+\frac{2m}{3}a_{abc}\nabla_b Y_c+ \frac{m}{3}(\nabla_b a_{abc})Y_c-\frac{m^2}{9} a_{abc}a_{bcd}Y_d=T_bT_bY_a.
\eeq
Using  the octonionic identities 
$\nabla_a a_{bcd}= mc_{abcd}$ and $a_{abe}a^{cd}{}_e=2\de_{ab}^{cd}+c_{ab}{}^{cd}$ (see the Appendix)
 together with  $T_aT_a=-(C_G-C_H)$, the $G/H$ equation above simplifies to
\beq
-\Box Y_a - \frac{2m}{3}a_{abc}\nabla_bY_c +6m^2Y_a  =C_GY_a.
\eeq
Here we  have also used the fact that the irrep ${\bf 7}$ of $SO(7)$ splits into $(1,1)\oplus (0,2)$ of $H=Sp_1^A\times Sp_1^{B+C}$ and that $C_H=\tfrac{12}{5}$ for both of these $H$ irreps.
So eliminating the $\Box$ term from the $\Delta_1$ and $G/H$ equations above gives 
\beq
	(\kappa_1^2-C_G)Y_a = \frac{2m}{3}a_{abc}\nabla _b Y_c.
\eeq

In order to extract the eigenvalues $\kappa_1^2$ from  this equation we  will have to square it. Let us define the operator 
\beq
DY_a\equiv a_{abc}\nabla _b Y_c.
\eeq
Taking the square of $D$ then goes as follows:
\beq
D^2Y_a=a_{abc}\nabla _b a_{cde}\nabla_dY_e=a_{abc}(\nabla _b a_{cde})\nabla_dY_e+a_{abc} a_{cde}\nabla _b\nabla_dY_e.
\eeq
Using identities from the Appendix this equation becomes 
\beq
 D^2Y_a= ma_{abc}c_{bcde}\nabla_dY_e+(2\de_{ab}^{de}+c_{ab}{}^{de})\nabla _b\nabla_dY_e= 4mDY_a+(6m^2-\Box)Y_a.
 \eeq
 Here we have also used the Ricci identity on 1-forms $[\nabla_a,\nabla_b]Y_c=R_{abc}{}^dY_d$ and the fact
 that the Riemann tensor term in the computation of $D^2$ above drops due to its contraction with $c_{eabc}$. Then since  
 the last term in the equation for $D^2Y_a$ above is just $\Delta_1$, the equation can be written 
 \beq
 D^2Y_a- 4mDY_a-\kappa_1^2Y_a=0.
 \eeq
If we now use $DY_a=\frac{3}{2m}(\kappa_1^2-C_G)Y_a$ and  $D^2Y_a=(\frac{3}{2m}(\kappa_1^2-C_G))^2Y_a$   we get the final result for the 1-form eigenvalues
 \beq
 \Delta_1=\frac{m^2}{9}(20C_G+14\pm2\sqrt{20C_G+49})=\frac{m^2}{9}(\sqrt{20C_G+49}\pm1  )^2-4m^2.
 \eeq
 The last form of the answer will be useful later.

\subsection{Spin 1/2 by squaring}
Before turning our attention to forms of rank two and three, and after that second rank symmetric tensors, we will check that our methods are able to produce the known result for the Dirac operator
acting on spin 1/2 modes. The spectrum of $\slashed D_{1/2} \equiv -i \slashed \nabla$ was derived in \cite{Nilsson:1983ru} by a different method based on the fibre bundle description of the
squashed seven-sphere. The virtue of the method in \cite{Nilsson:1983ru} is that one can follow the eigenvalues as one turns the squashing parameter from the round sphere value to
the squashed Einstein space value.  This nice feature is unfortunately lacking for the methods used in this paper.

To apply our present techniques we start by squaring the Dirac operator in order to obtain a situation that is similar to the one for the Hodge-de Rham operators on $p$-forms. Using
$\Gamma^a\nabla_a\Gamma^b\nabla_b\psi=(\Box+\thalf\Gamma^{ab}[\nabla_a,\nabla_b])\psi$ we find
\beq
-i\slashed \nabla \psi=\lambda\psi \Rightarrow (-\Box+\frac{R}{4})\psi=\lambda^2\psi.
\eeq
It is now possible to use the $G_2$ structure to split the tangent space spinor irrep into $G_2$ irreps as ${\bf 8}\rightarrow {\bf 7}\oplus {\bf 1}$, or in terms of indices $A=(a,8)$. 
Then by defining  two spinors $\eta$ and $ \eta_a$ we can expand a general Dirac spinor as follows
\beq
\psi=V^a\eta_a+f\,\eta,
\eeq
where $V^a$ is a vector field and $f$ a scalar field on the seven-sphere. While $\eta$ is the standard Killing spinor with components $\eta_B=\de_{B}^8$, $\eta_a$ is defined by $\eta_a\equiv -i\Gamma_a \eta$
which  implies the its explicit form is  $(\eta_a)_B = \delta_{aB}$.
These spinors are
linearly independent and satisfy
\beq
\nabla_a\eta=\tfrac{m}{2}\eta_a, \,\,\,\,\,\nabla_a\eta_b=-\tfrac{m}{2}\de_{ab}\eta+\tfrac{ m}{2}a_{abc}\eta_c.
\eeq
These equations imply
\beq
\Box\eta=-\tfrac{7m^2}{4}\eta,\,\,\,\,\Box\eta_a=-\tfrac{7m^2}{4}\eta_a,
\eeq
from which we obtain the equations
\beqa\label{BoxSpinor}
\Box(f\eta)&=&(\Box f-\tfrac{7m^2}{4}f)\eta+m(\nabla^af)\eta_a,\\ 
\,\,\,\,\Box(V^a\eta_a)&=&(\Box V^a-\tfrac{7m^2}{4}V^a)\eta_a+m(\nabla^aV^b)a_{abc}\eta_c.
\eeqa

Consider now the coset master equation for Dirac spinor modes. Squaring it gives
\beq
\Box \psi-\tfrac{m}{6}a_{abc}\Gamma_{ab}\nabla_c\psi-\tfrac{7m^2}{12}\psi+\tfrac{m^2}{144}c_{abcd}\Gamma_{abcd}\psi=T_aT_a\psi.
\eeq
So eliminating the $\Box$ term and inserting $R=42m^2$ we find that the equation we need to solve reads
\beq
\label{diracboxeq}
-\lambda^2\psi-\tfrac{m}{6}a_{abc}\Gamma_{ab}\nabla_c\psi+\tfrac{119m^2}{12}\psi+\tfrac{m^2}{144}c_{abcd}\Gamma_{abcd}\psi=T_aT_a\psi.
\eeq
In order to get the last term on the LHS in a nice form we introduce the projection operators for the $G_2$ split $\psi=\psi_1+\psi_7$ corresponding to  ${\bf 8}\rightarrow {\bf 1}\oplus {\bf 7}$. 
They are (see Appendix for more details)
\beq
P^s_1=\tfrac{1}{8}(1-\tfrac{1}{24}c_{abcd}\Gamma_{abcd}),\,\,\,\,P^s_7=1-P_1=\tfrac{1}{8}(7+\tfrac{1}{24}c_{abcd}\Gamma_{abcd}).
\eeq
In terms of these projectors the sum of the last two terms on the LHS above becomes
\beq
\tfrac{119}{12}\psi+\tfrac{1}{144}c_{abcd}\Gamma_{abcd}\psi=\tfrac{35}{4}P^s_1\psi+\tfrac{121}{12}P^s_7\psi.
\eeq
Then using the octonionic version of the seven-dimensional gamma matrices given in the Appendix, some algebra gives
\beq
a_{abc}\Gamma_{ab}\nabla_c\psi=(-6\nabla_aV^a-21mf)\eta+(2a^{abc}\nabla_a V_b-9mV^c+6(\nabla^cf))\eta_c.
\eeq

Inserting these results into (\ref{diracboxeq}) we get an equation that can be hit by the projectors $P_1^s$ and $P_7^s$ leading to the two equations
\beq
\lambda^2 f=\tfrac{20}{9}m^2 C_G f+\tfrac{35}{4}m^2f+m(\nabla_aV^a+\tfrac{7}{2}mf),
\eeq
and 
\beq
\lambda^2 V_a=\tfrac{20}{9}m^2 (C_G -\tfrac{12}{5})V_a+(\tfrac{121}{12}+\tfrac{3}{2})m^2V_a-\tfrac{1}{3}ma_{abc}\nabla_bV_c - m\nabla_a f.
\eeq
To solve these equations we either have $f=0$ or $f \neq 0$. In the former case the first equation gives $\nabla^aV_a=0$ which implies that the second equation for $V_a$ can be
analysed in exactly the same way as for the 1-forms discussed previously.
Not surprisingly the result is also of $\sqrt{49}$-form and reads
\beq
\lambda^2=m^2(\tfrac{1}{6}\pm\tfrac{1}{3}\sqrt{20C_G+49})^2.
\eeq

Turning to the latter case, that is $f\neq 0$, we start by taking the divergence of the second equation and use the fact that $\Box f=-\frac{20m^2}{9}C_Gf$ to find
\beq
(\tfrac{20}{9}m^2C_G+\tfrac{25}{4}m^2-\lambda^2)(\nabla_aV^a)=-\tfrac{20}{9}m^3C_G\,f.
\eeq 
Now the system of two equations for the functions $f$ and $\nabla^aV_a$ has the solutions
\beq
\lambda^2=m^2(\thalf\pm\tfrac{1}{3}\sqrt{20C_G+81})^2.
\eeq

These two sets of eigenvalues are consistent with the known result from \cite{Nilsson:1983ru}. However, in that paper there is no sign ambiguity
since there one obtains $\lambda$ instead of $\lambda^2$. This problem is easily eliminated by applying our present  result to the spin $2^+$ 
supermultiplet. 

This completes the review. We now turn to 2-forms where the results are new.
\subsection{2-forms}

In terms of the Hodge-de Rham operator on 2-forms, the equation to be solved is
\beq
\Delta_2 Y_{ab}=-\Box Y_{ab}-2R_{acbd}Y^{cd}-2R_{[a}{}^cY_{b]c}=\kappa_2^2 Y_{ab}.
\eeq
Using \eqref{eq:RiemannDef} for the Riemann tensor gives the following equation for the box operator
\beq
\label{deltatwo}
\Delta_2:\,\,\,\,\Box Y_{ab}=-\kappa^2_2 Y_{ab}-2W_{acbd} Y^{cd}+10m^2 Y_{ab}.
\eeq

As for the previous cases the next step is to express the box operator in terms of algebraic objects on the coset manifold. The master coset formula (\ref{cma})
for 2-forms reads
\beq
\nabla_a Y_{bc}=-T_aY_{bc}-f_{a}{}^{d}{}_{[b}Y_{c]d},
\eeq 
which  can be written
\beq
\tilde\nabla_a Y_{bc}=-T_aY_{bc},
\eeq
by indroducing the "$G_2$-derivative" $\tilde\nabla_a$  defined on 2-forms by
\beq
\tilde\nabla_a Y_{bc}\equiv \nabla_a Y_{bc}+\frac{2m}{3}a_{a[b}{}^d Y_{c]d}.
\eeq
This derivative is "$G_2$" in the sense that it satisfies $\tilde\nabla_a a_{bcd}=0$ and $\tilde\nabla_a c_{bcde}=0$. This step is not crucial here but we will find more implications later of the presence of the 
$G_2$ holonomy so we will have reason to return to this derivative then. (For more details about $G_2$ in this context, see for instance \cite{House:2004pm}.)
To get an algebraic expression for the box operator we can now square the master coset equation, $\tilde\nabla_a Y_{bc}=-T_aY_{bc}$, to get $\tilde\Box Y_{ab}=T_cT_cY_{ab}$. This equation
can also be written as
\beq
\label{goverhtwoforms}
G/H:\,\,\,\,\,\tilde\Box Y_{ab}=(\Box -\frac{10}{9}m^2)Y_{ab}+\frac{4}{3}ma_{cd[a}\nabla_{|c}Y_{d|b]}-\frac{2}{9}m^2c_{abcd}Y_{cd}=T_cT_cY_{ab}.
\eeq

Finally, to get the equation that  needs to be solved to find the eigenvalues we just eliminate the box operator from (\ref{deltatwo})  and (\ref{goverhtwoforms}). Using the projectors defined in the Appendix
gives the following useful form of the resulting equation 
\beq\label{algebraic2form}
(\kappa^2_2-12m^2+T_cT_c)Y_{ab}=-W_{abcd}Y^{cd}-\frac{4}{3}m^2(P_{14}Y)_{ab}+\frac{4}{3}m(a_{cd[a}\tilde\nabla_{|c}Y_{d|b]}),
\eeq
which, again, is not entirely algebraic due to the appearance of the operator $\tilde\nabla_a$ in the last term. This also happened in the case of the
1-form where it was easy to handle by a squaring procedure. This trick is a bit harder to apply in the case of 2-forms (as well as for the other
operators to be discussed later) as will become clear shortly. Compared to the 1-form case there is also a new feature here namely the
presence of the Weyl tensor in one of the terms which will  cause additional complications. The approach used in this paper to deal with the Weyl tensor terms, in this and the other cases discussed below,
 is explained in the Appendix.

Thus there are two new issues when trying to solve the   2-form equation (\ref{algebraic2form}): The Weyl tensor term and the $\tilde\nabla$ term which now tends to lead to symmetrised derivatives when squared. 
To deal with the former issue we recall how $T_aT_a$ can be expressed in terms of the  Casimir operators for the groups involved: $T_aT_a=-(C_G-C_H)$.
This, however, has implications for the how the spectrum is organised in terms of towers. To see this we  use the tangent space decomposition
$
SO(7)\rightarrow G_2 \rightarrow Sp^A_1\times Sp^{B+C}_1
$
which makes it possible to read off   the relevant decompositions directly from the McKay and Patera tables \cite{Mckay:1981mp}. 
%
In the case of the 2-form the composition reads, see \cite{Duff:1986hr} or the summary in \cite{Nilsson:2018lof},
\beq
\bf{21}\rightarrow \bf{7}\oplus \bf{14} \rightarrow (\bf{(1,1)}\oplus \bf{(0,2)})\oplus  ((0,2)\oplus (1,3) \oplus (2,0)).
\eeq
This means that the towers will be tabulated according to their $H$ irrep which will thus obscure the  $G_2$ structure of the spectrum. As is explained in the Appendix
this problem is  automatically eliminated once the Weyl tensor term is analysed and the result combined with the result from the $C_H$ term. In fact, adding these two terms  gives
$\tfrac{24}{5}Y_{ab}=\tfrac{32}{3}m^2Y_{ab}$ for all the $H$ irreps in the $G_2$ irrep ${\bf 14}$. Recall that the corresponding answer for the irrep ${\bf 7}$ is $\tfrac{12}{5}=\tfrac{16}{3}m^2$.

Using  this insight from the Appendix, namely   that the sum $C_H+Weyl$ has a common value on all $H$ irreps in the decomposition of each $G_2$ irrep arising from the $SO(7)$ representation ${\bf 21}$, 
it becomes possible  to split the 2-form equation into two by acting on it with the projectors $P_7$ and $P_{14}$ and insert the respective values of $C_H+Weyl$. 
We find, using the definitions $Y^{(7)}_{ab}\equiv (P_7Y)_{ab}$ and $Y_a\equiv a_{abc}Y_{bc}^{(7)}$,
\beq
{\bf 7}:\,\,\,\,\,\,\kappa_2^2Y_a=C_GY_{a}+(12+\frac{4}{3}-\frac{16}{3})m^2Y_a-\frac{4m}{3}(a_{abc}\nabla_bY_c),
\eeq
\beq
{\bf 14}:\,\,\,\,\,\,\kappa_2^2Y^{(14)}_{ab}=C_GY^{(14)}_{ab}+(12-\frac{4}{3}-\frac{32}{3})m^2Y^{(14)}_{ab}+\frac{2m}{3}(P_{14})_{ab}^{cd}(\nabla_c Y_{d}).
\eeq
Note that the last term in the second equation contains $Y_a$ and thus seems to mix with the first equation. However, the structure of the derivative terms as a 2 by 2 matrix 
shows that this term will have no role in determining the eigenvalues for the 2-form as will be clear below. To proceed we write these two equations as
\beq
{\bf 7}:\,\,\,\,\,\,\kappa_2^2Y_a=\frac{m^2}{9}(20C_G+72)Y_a-\frac{4}{3}m(a_{abc}\nabla_bY_c),
\eeq
\beq
{\bf 14}:\,\,\,\,\,\,\kappa_2^2Y^{(14)}_{ab}=\frac{m^2}{9}\,20C_GY^{(14)}_{ab}+\frac{2}{3}m(P_{14}\nabla Y)_{ab}.
\eeq

Consider first the possibility to take the square of the first equation above, the one for $Y_a$. However, although this will give rise to a calculation quite similar to the one for the 1-form, there is one important difference.
While the 1-form is transverse (i.e., divergence free) this is not the case for $Y_a$ coming from the 2-form. Therefore one needs to analyse the two equations for the 2-form in stages:\\
\\
1) Either $Y_a=0$ or $Y_a\neq 0$, 2) when $Y_a\neq 0$ then either $\nabla_aY_a=0$ or $\nabla_a Y_a\neq 0$.\\
\\
So, when $Y_a=0$ the equation left to solve is the ${\bf 14}$-equation with the derivative term set to zero. This equation therefore gives the eigenvalues
\beq
\Delta_2=\frac{m^2}{9}\,20\,C_G.
\eeq
According to the characterisation mention above, this result leads to an energy $E_0$ of the $\sqrt{9}$-form. This  fits nicely with a supermultiplet containing spins $(1^+, 1/2, 1/2, 0^+)$ with masses related to the operators
$(\Delta_2, {\slashed D}_{3/2}, {\slashed D}_{3/2}, \Delta_L)$ since we know from Ref. [198] in \cite{Duff:1986hr}  that $\Delta_L$ also has an eigenvalue leading to 
a $\sqrt{9}$-form. One can check that the corresponding  values of $E_0$ work out as they should in relation to supersymmetry. 
This eigenvalue of $\Delta_L$ leading to the $\sqrt{9}$-form will be derived in detail  later in this section.

We now turn to the cases with $Y_a\neq 0$, namely $\nabla_aY_a=0$ and $\nabla_a Y_a\neq 0$. Clearly, when $\nabla_aY_a=0$ the calculation will follow the one for 1-forms very closely. 
Indeed, the result is also of  $\sqrt{49}$-form and reads
\beq
\Delta_2=\frac{m^2}{9}(\sqrt{20 C_G+49}\pm 2)^2-m^2.
\eeq
Although being of $\sqrt{49}$-form, this 2-form eigenvalue formula differs from the one for 1-forms given  in \cite{Duff:1986hr}, and also derived above,
 \beq
 \Delta_1=\frac{m^2}{9}(20 C_G+ 14\pm 2 \sqrt{20 C_G + 49})=\frac{m^2}{9}(\sqrt{20 C_G+49}\pm 1)^2-4m^2.
\eeq
The differences are partly compensated for by the
different relations between the Hodge-de Rham operators and the respective $M^2$ operators \cite{Duff:1986hr}:
\beq
M^2(1^+)=\Delta_2,
\eeq
\beq
M^2(1^-)=\Delta_1+12m^2\pm6m\sqrt{\Delta_1+4m^2}=(\sqrt{\Delta_1+4m^2}\pm 3m)^2-m^2.
\eeq
If the $\Delta_1$ eigenvalues are inserted into this formula for  $M^2(1^-)$ we see that the last term, that is $-4m^2$ in $ \Delta_1$, is really required to have a chance for supersymmetry to work
in a supermultiplet containing both $1^+$ and $1^-$ fields. In fact, the field content of the spin 3/2 supermultiplets has this property and requires
  $E_0(1^+)=E_0(1^-)\pm 1$. To verify this relation we need the energy formula for spin 1 unitary $SO(3,2)$ irreps
\beq
E_0(1^{\pm})=\frac{3}{2}+\half\sqrt{\frac{M^2}{m^2}+1}.
\eeq
This expression gives the energy values
\beq
E_0(1^{-(1),(2)})=\frac{3}{2}\pm\frac{3}{2}\mp\frac{1}{6}+\frac{1}{6}\sqrt{20C_G+49},
\eeq
where the first $\pm$ sign refers to the two towers of spin $1^{-(1),(2)}$ fields in $AdS_4$ supergravity theory. 
Comparing these to 
\beq
E_0(1^+)=\frac{3}{2}\pm\frac{1}{3}+\frac{1}{6}\sqrt{20C_G+49},
\eeq
one finds (using also the spectrum of the spin 3/2 fields in these supermultiplets) that  it is possible to eliminate the sign ambiguities in the spectra of $\Delta_1$ and $\Delta_2$. For more details see \cite{Nilsson:2021kn}.

Finally, when $\nabla_a Y_a\neq 0$ we take the divergence of the ${\bf 7}$ projected part of the equation and use $a_{abc}\nabla_a\nabla_bY_c=0$ to obtain a simple equation
giving the eigenvalue
\beq
\kappa^2_2=\frac{m^2}{9}(20C_G+72)
\eeq
This completes the analysis of the 2-form eigenvalue spectrum. A  list of the obtained eigenvalues can be found in the Conclusions.
\subsection{3-forms} 
We now turn to 3-forms leaving the discussion of  the Lichnerowicz modes to  the next subsection. The reason for doing the analysis in this order is that
we will obtain some equations in the 3-form case that can be applied also  for Lichnerowicz. 

The  equation for 3-forms that we wish to solve reads
\beq
\Delta_3Y_{abc}=-\Box Y_{abc}+6R^d{}_{[ab}{}^eY_{c]de}+3R_{[a}{}^dY_{bc]d}=\kappa_3^2Y_{abc}.
\eeq
Inserting the Riemann and Ricci tensors of the squashed seven-sphere this equation becomes
\beq
\Delta_3:\,\,\,\,\,\Box Y_{abc}=-\kappa_3^2Y_{abc}+6W^d{}_{[ab}{}^eY_{c]de}+12m^2Y_{abc}.
\eeq
Squaring the coset master equation for 3-forms  gives
\beq
G/H:\,\,\,\,\,-(C_G-C_H)Y_{abc}=\Box Y_{abc}-2ma_{d[a}{}^e\tilde\nabla_{|d|}Y_{bc]e}+\frac{m^2}{3}(4Y_{abc}+2c_{[ab}{}^{de}Y_{c]de}),
\eeq
where for later convenience we have used the $G_2$ derivative
\beq
\tilde\nabla_d Y_{abc}\equiv \nabla_d Y_{abc}-ma_{d[a}{}^eY_{bc]e}.
\eeq
Note that this implies that the divergence
\beq
\tilde\nabla^a Y_{abc}=\nabla^a Y_{abc}-\tfrac{m}{3}(a_{aa}{}^eY_{bce}+2a_{a[b}{}^eY_{c]ae})=\tfrac{2m}{3}a_{[b}{}^{de}Y_{c]de}.
\eeq
Eliminating the box-operator term from the above equations gives
\beq\label{Algebraic3Form}
\kappa_3^2Y_{abc}=C_GY_{abc}-(C_HY_{abc}-6W^d{}_{[ab}{}^eY_{c]de})+\frac{m^2}{3}(40Y_{abc}+2c_{[ab}{}^{de}Y_{c]de})-2ma_{d[a}{}^e\tilde\nabla^dY_{bc]e}.
\eeq
To proceed we decompose the 3-form into $G_2$ irreps as ${\bf 35}={\bf 1}\oplus{\bf 7}\oplus {\bf 27}$. By defining $Y \equiv a_{abc}Y_{abc}$ and $Y_{a} \equiv c_{abcd}Y_{bcd}$ we can split the 3-form as follows
\beq
Y_{abc}=Y^{(1)}_{abc}+Y^{(7)}_{abc}+Y^{(27)}_{abc}=\frac{1}{42}a_{abc}Y-\frac{1}{24}c_{abcd}Y_d+(P_{27}Y)_{abc},
\eeq
where we have utilised the 3-form projectors defined in the Appendix. The term $C_HY_{abc}-6W^d{}_{[ab}{}^eY_{c]de}$ can be written (see Appendix)
\beq\label{Weyl3Form}
C_HY_{abc}-6W^d{}_{[ab}{}^eY_{c]de}=-\frac{12}{5}(\frac{1}{24}c_{abcd}Y_d)+\frac{28}{5}(P_{27}Y)_{abc}.
\eeq
Apart from this there are two more problematic terms, the $c_{abcd}$-term and the derivative term that will force us to square also this  equation.

However, before doing that we will follow the strategy in the 2-form case and  start by splitting the equation into $G_2$ pieces. The singlet term ${\bf 1}$ is obtained by contracting all three indices  by $a_{abc}$ which gives
\beq
\kappa_3^2Y=C_GY+\frac{m^2}{3}(40Y+2a_{abc}c_{[ab}{}^{de}Y_{c]de})-2ma_{abc}a_{d[a}{}^e\tilde\nabla^dY_{bc]e}.
\eeq
Cleaning up the structure constant factors this equation becomes
\beq
{\bf 1}:\,\,\,\,\,\,\kappa_3^2Y=(C_G+16m^2)Y+2m\tilde\nabla_aY_a.
\eeq
Turning to the ${\bf 7}$ part, we find after some algebra
\beq
{\bf 7}:\,\,\,\,\,\,\kappa_3^2Y_a=(C_G+6m^2)Y_a-\frac{12m}{7}\nabla_aY+\frac{m}{3}a_{abc}\nabla_bY_c+2ma_{bcd}\nabla_bY^{(27)}_{cda}.
\eeq

When we now come to the last part, the ${\bf 27}$, it will require some new steps that will  relate it to the metric and the Lichnerowicz equation.
This connection is in fact already indicated by the last term in the ${\bf 7}$ equation just discussed: $a_{bcd}\nabla_bY^{(27)}_{cda}$. This expression suggests that 
the 2-index tensor $a_{acd}Y^{(27)}_{bcd}$ will be useful as we now elaborate upon (see \cite{House:2004pm} for a closely related discussion). Let us define a 2-index tensor from the ${\bf 27}$ part of
the 3-form by
\beq
\tilde Y_{ab}\equiv a_{acd}Y^{(27)}_{bcd},
\eeq
where we use a tilde to avoid confusing it with the antisymmetric 2-form discussed previously.
Clearly, the symmetric and traceless part of $\tilde Y_{ab}$ is also in the irrep ${\bf 27}$. However, a nice and indeed very useful fact about this definition
of $\tilde Y_{ab}$ is that it is automatically symmetric and traceless. The tracelessness  follows immediately from the identity $a_{abc}c_{abcd}=0$, 
or, using the 3-form projectors in the Appendix, from $P_1P_{27}=0$,  while the vanishing of its 
antisymmetric part follows from $a_{abc}\tilde Y_{bc}=0$ and $c_{abcd}\tilde Y_{cd}=0$\footnote{Note that these two conditions can be combined to the identity since 
 $\de_{ab}^{cd}=\thalf(a_{abe}a^{cde}-c_{ab}{}^{cd})$.}. These two results can be checked as follows:
\beq
a_{abc}\tilde Y_{bc}=a_{abc}a_{bde}Y^{(27)}_{cde}=-(2\de_{ac}^{de}+c_{acde})Y^{(27)}_{cde}=-c_{acde}Y^{(27)}_{cde}=0,
\eeq
where we have used $P_7P_{27}=0$ in the last step, and 
\beq
c_{abcd}\tilde Y_{cd}=c_{abcd}a_{cef}Y^{(27)}_{def}=-6a_{[ab}{}^{[e}\de_{d]}^{f]}Y^{(27)}_{def}=4a_{[a}{}^{de}Y^{(27)}_{b]de}=0.
\eeq
Here the very last equality is a consequence of the identity
\beq
a_{abc}(P_{27})_{bcd}{}^{efg}=a_{(a}{}^{[ef}\de_{d)}^{g]}-\frac{1}{7}\de_{ad}a^{efg}.
\eeq
Clearly, also $a_{abc}(P_{27})_{abc}{}^{efg}=0$ again showing that the trace of $\tilde Y_{ab}$ vanishes.

An alternative, and perhaps more direct, way to see that $\tilde Y_{ab}$ is symmetric and traceless is to just insert the expression for the 
$P_{27}$ projector which gives
\beq
	a_{abc}Y^{(27)}_{bcd} = \frac{1}{2}(a_{abc}Y_{bcd}+a_{dbc}Y_{bca})-\frac{1}{7}\delta_{ad}Y. 
\eeq

Finally, for the definition $\tilde Y_{ab}\equiv a_{acd}Y^{(27)}_{bcd}$ to be useful we need to give also the inverse 
relation\footnote{Here one may use the fact that $Y^{(27)}_{abc}=(P_{27}Y^{(27)})_{abc}$ implies the identity $Y^{(27)}_{abc}=-\tfrac{3}{2}c_{de[ab}Y_{c]de}^{(27)}$.}:
\beq
Y^{(27)}_{abc}=\frac{3}{4}a_{d[ab}\tilde Y_{c]d}.
\eeq
The decomposition of the 3-form therefore takes the following simple form in terms of $Y, Y_a, \tilde Y_{ab}$, respectively in irreps ${\bf 1}, {\bf 7}, {\bf 27} $, 
\beq
Y_{abc}=\frac{1}{42}a_{abc}Y-\frac{1}{24}c_{abcd}Y_d+\frac{3}{4}a_{d[ab}\tilde Y_{c]d}.
\eeq

With these preliminary results at hand it is now a rather straightforward matter to project the equation \eqref{Algebraic3Form} onto its ${\bf 27}$ component.  One finds
\beq
\kappa_3^2\tilde Y_{ab}=(C_G-\frac{m^2}{3}(40-\frac{4}{3}))\tilde Y_{ab}-(C_H\de_{(ab)}^{cd}+2W_{acbd})\tilde Y_{cd}-\frac{2}{3}ma_{cd(a}\tilde\nabla^c\tilde Y_{b)}{}^d-\frac{m}{3}(\tilde\nabla_{(a}Y_{b)}-\frac{1}{7}\de_{ab}\tilde\nabla^cY_c).
\eeq
Again we can replace the second bracket by its common eigenvalue on all $H$ irreps in the $G_2$ irrep ${\bf 27}$ as explained in the Appendix, that is with  $\tfrac{28}{5}=\tfrac{112m^2}{9}$.

We now insert this into the ${\bf 27}$ equation and sum up what we have found so far:
\beq
{\bf 1}:\,\,\,\,\kappa_3^2Y=(C_G+16m^2)Y+2m\tilde\nabla_aY_a,
\eeq
\beq
{\bf 7}:\,\,\,\,\kappa_3^2Y_a=(C_G+6m^2)Y_a-\frac{12m}{7}\nabla_aY+\frac{m}{3}a_{abc}\nabla_bY_c+2ma_{bcd}\nabla_bY^{(27)}_{cda}.
\eeq
\beq
{\bf 27}:\,\,\,\,\kappa_3^2\tilde Y_{ab}=(C_G+\frac{4m^2}{9})\tilde Y_{ab}-\frac{2}{3}ma_{cd(a}\tilde\nabla^c\tilde Y_{b)}{}^d-\frac{m}{3}(\tilde\nabla_{(a}Y_{b)}-\frac{1}{7}\de_{ab}\tilde\nabla^cY_c).
\eeq

Note that the term $2ma_{bcd}\nabla_bY^{(27)}_{cda}$ appearing  in the ${\bf 7}$ equation  can be replaced by terms containing only $Y$ and $Y_a$. This relation is derived from
the gauge condition  $\nabla^aY_{abc}=0$ as follows:
\beq
\nabla^a(Y^1_{abc}+Y_{abc}^7+Y_{abc}^{27})=0.
\eeq
Contracting this equation with $a_{dbc}$ and using $\nabla_a a_{bcd}=mc_{abcd}$ we get
\beq
\nabla^a a_{dbc}(Y^1_{abc}+Y_{abc}^7+Y_{abc}^{27})-mc_{adbc}(Y^1_{abc}+Y_{abc}^7+Y_{abc}^{27})=0.
\eeq
However, since $c_{abcd}$ projects onto $Y_a$ we can use the relations
\beq
a_{dbc}Y_{abc}^1=\frac{1}{42}a_{dbc}a_{abc}Y=\frac{1}{7}\de_{da}Y,\,\,\,\,a_{dbc}Y_{abc}^7=-\frac{1}{24}a_{dbc}c_{abce}Y_e=-\frac{1}{6}a_{dab}Y_b, \,\,\,a_{dbc}Y_{abc}^{27}=\tilde Y_{da},
\eeq
to rewrite the above 3-form gauge condition   as
\beq
\nabla^a\tilde Y_{ab}+\frac{1}{7}\nabla_bY-\frac{1}{6}a_{bcd}\nabla_cY_d+mY_b = 0.
\eeq
This equation will be used when we summarise the system of equations to be solved below.

Finally, we show that the last bracket in the ${\bf 27}$ equation will not play a role in the analysis of the spectrum. This can be seen by writing the three equations above in matrix form as
\beq
\kappa_3^2 
\begin{pmatrix}
Y^1 \\
Y^7 \\
Y^{27}
\end{pmatrix}
=
\begin{pmatrix}
A & B & 0\\
C & D & 0\\
0 &E & F
\end{pmatrix}
\begin{pmatrix}
Y^1 \\
Y^7 \\
Y^{27}
\end{pmatrix}
\eeq
The eigenvalues must be the roots of  the equation $\det(X-\kappa_3^2{\bf 1})=0$, where $X$ is the matrix appearing above. Since the element $E$ does not enter this equation
we can proceed and solve the three equations in two steps, first the two coupled equations for $Y$ and $Y_a$
\beq
{\bf 1}:\,\,\,\,\kappa_3^2Y=(C_G+16m^2)Y+2m\nabla_aY_a,
\eeq
\beq
{\bf 7}:\,\,\,\,\kappa_3^2Y_a=(C_G+4m^2)Y_a-2m\nabla_aY+\frac{2m}{3}a_{abc}\nabla_bY_c,
\eeq
and then the single equation involving only $\tilde Y_{ab}$
\beq
{\bf 27}:\,\,\,\,\kappa_3^2\tilde Y_{ab}=(C_G+\frac{4m^2}{9})\tilde Y_{ab}-\frac{2m}{3}a_{cd(a}\tilde\nabla^c\tilde Y_{b)}{}^d.
\eeq

We start by solving the first two coupled equations. There are two distinct cases: either 1) $Y=0$ or 2) $Y\neq 0$.

Case 1): Setting $Y=0$ in the ${\bf 1}$ equation gives $\nabla_aY_a=0$ which implies that the ${\bf 7}$ equation is of exactly of the same type as the equation solved
previously for 1-forms, but with different coefficients. However, one has to pay attention to the fact that the eigenvalues dealt with here are for 3-forms, not 1-forms,
so solving the present equation for $Y_a$ involves some new steps. This fact will become clear directly when squaring the operator $(DY)_a\equiv a_{abc}\nabla_bY_c$:
\beqa
(DDY)_a:&=&a_{abc}\nabla_ba_{cde}\nabla_dY_e=a_{abc}a_{cde}\nabla_b\nabla_dY_e+ma_{abc}c_{bcde}\nabla_dY_e=\cr
&&(2\de_{ab}^{de}+c_{abde})\nabla_b\nabla_dY_e+4ma_{ade}\nabla_dY_e.
\eeqa
Since the $c$-term vanishes this equation simplifies to
\beq
(DDY)_a=4m(DY)_a-\Box Y_a+6m^2Y_a.
\eeq
At this point the $\Box Y_a$ term must be related to the 3-form eigenvalue $\kappa_3^2$. Using $Y=0$ and $\nabla^aY_{abc}=0$, we find that
\beqa
\Box Y_a&=&\Box c_{abcd}Y_{bcd}=c_{abcd}\Box Y_{bcd}-2m^2Y_a=\cr
&&c_{abcd}(-\kappa_3^2Y_{bcd}+6W^e{}_{[bc}{}^fY_{d]ef}+12m^2Y_{bcd})-2mY_a=-(\kappa_3^2-10m^2)Y_a,
\eeqa
where we have also used $c_{abcd}W^e{}_{[bc}{}^fY_{d]ef}=W_{adef}Y_{def}=0$ (where the first equality follows from $P_7^{(2)}W=0$). The $(DDY)_a$ equation above then reads 
\beq
(DDY)_a=4m(DY)_a+(\kappa_3^2-4m^2)Y_a.
\eeq
Inserting the expression for $(DY)_a$ from the eigenvalue equation ${\bf 7}$ above we find
\beq
(\kappa_3^2-C_G-4m^2)^2-\frac{8m^2}{3}(\kappa_3^2-C_G-4m^2)-\frac{4m^2}{9}(\kappa_3^2-4m^2)=0,
\eeq
which has the following two solutions (inserting $1=\tfrac{20}{9}m^2$ in front of $C_G$)
\beq
\Delta_3=\frac{m^2}{9}(\sqrt{20C_G+49}\pm 1)^2.
\eeq
These eigenvalues are obtained for $\Delta_3=Q^2$ so it is gratifying to find its eigenvalues come out as a square.

Case 2): Now we turn to the second case where $Y\neq 0$. Then either $\nabla_aY_a=0$ or $\nabla_aY_a\neq 0$. The former case gives a simple
equation for the eigenvalue directly from the ${\bf 1}$ equation. However, taking the divergence of the ${\bf 7}$ equation tells us that $\Box Y=0$ so the 
${\bf 1}$ equation reduces to 
\beq
\kappa_3^2=16m^2,
\eeq
which is the eigenvalue of the singlet constant mode $Y_{abc}=a_{abc}$ as is easily verified. So let us turn to the latter case with $\nabla_aY_a\neq 0$. Taking 
the divergence of the ${\bf 7}$ equation gives 
\beq
\Box Y=\frac{1}{2m}(-\kappa_3^2+C_G+4m^2)\nabla_aY_a.
\eeq
 Inserting this equation back into the ${\bf 1}$ equation, recalling that $\Box Y = -C_GY$,  gives
\beq
(\kappa_3^2-C_G-16m^2)(\kappa_3^2-C_G-4m^2)=\frac{9}{5}C_G.
\eeq
 This equation has the following solutions (inserting again $1=\tfrac{20}{9}m^2$ in front of $C_G$):
 \beq
 \Delta_3=\tfrac{m^2}{9}(\sqrt{20C_G+81}\pm 3)^2.
 \eeq 
 Note that the single eigenvalue found above, $\kappa_3^2=16m^2$, belongs to the plus branch. 
 
 We now turn to the ${\bf 27}$ part of the 3-form equation
 \beq\label{27KappaEquation}
{\bf 27}:\,\,\,\,\kappa_3^2\tilde Y_{ab}=(C_G+\frac{4m^2}{9})\tilde Y_{ab}+\frac{2m}{3}a_{c(a}{}^d\tilde\nabla^c\tilde Y_{|d|b)}.
\eeq
The derivative term requires as usual a squaring of the whole equation. The $G_2$ derivative on $\tilde Y_{ab}$ is given by
\beq
(D\tilde Y)_{ab}\equiv 2a_{cd(a}\tilde\nabla^c\tilde Y_{b)}{}^d=\tilde\nabla^c(a_{ca}{}^d\tilde Y_{db}+a_{cb}{}^d\tilde Y_{ad}),
\eeq
and computing its square, using the fact that $\tilde\nabla_a$ is zero on both $a_{abc}$ and $c_{abcd}$, gives
\beq
(DD\tilde Y)_{ab}\equiv \tilde\nabla^e\tilde\nabla^f(a_{ea}{}^ca_{fc}{}^d\tilde Y_{db}+a_{eb}{}^ca_{fc}{}^d\tilde Y_{ad}+a_{ea}{}^ca_{fb}{}^d\tilde Y_{cd}+a_{eb}{}^ca_{fa}{}^d\tilde Y_{cd}).
\eeq
Clearly,  the last two terms in this expression contain a symmetric combination of the two $G_2$ covariant derivatives. Such terms can be a real obstacle
to carrying through the calculation but we will see below that there is a trick that can  be used to eliminate this issue. Before applying this trick we simplify the other terms to get
\beq
(DD\tilde Y)_{ab}=-2\tilde\Box \tilde Y_{ab}+2\tilde\nabla^c\tilde\nabla_{(a} \tilde Y_{b)c}-2c_{cde(a}\tilde\nabla^c\tilde\nabla^d\tilde Y_{b)}{}^e+2a_{e(a}{}^ca_{|f|b)}{}^d\tilde\nabla^e\tilde\nabla^f\tilde Y_{cd}.
\eeq

We now address the four terms on the RHS of this equation  starting with last one. The trick used to deal with the symmetric combination of derivatives in this term is to reintroduce
the 3-form via $Y^{(27)}_{abc}=\tfrac{3}{4}a_{d[ab}\tilde Y_{c]d}$ temporarily giving
\beq\label{SymmetricDerivative27}
a_{e(a}{}^ca_{|f|b)}{}^d\tilde\nabla^e\tilde\nabla^f\tilde Y_{cd}=a_{e(a}{}^c\tilde\nabla^e\tilde\nabla^fa_{|f|b)}{}^d\tilde Y_{cd}=a_{e(a}{}^c\tilde\nabla^e\tilde\nabla^f(4Y^{(27)}_{|f|b)c}-a_{b)c}{}^d\tilde Y_{fd}-a_{|cf|}{}^d\tilde Y_{b)d}).
\eeq
The first term in the last expression is just the $G_2$ covariant divergence of the 3-form which can be seen to satisfy
\beq
\tilde\nabla^aY_{abc}^{(27)}=\nabla^aY_{abc}^{(27)}=\nabla^aY_{abc}=0,
\eeq
where in the first equality we have used the fact that $\tilde Y_{ab}$ is symmetric and  in the second that  
 $\nabla^aY_{abc}^{(1)}=\nabla^aY_{abc}^{(7)}=0$. Thus $\nabla^aY_{abc}=0$ is implies  $\tilde\nabla^a\tilde Y_{ab}=0$ which gives the following much simpler form of the expression above
\beq
a_{e(a}{}^ca_{|f|b)}{}^d\tilde\nabla^e\tilde\nabla^f\tilde Y_{cd}=-a_{e(a}{}^c\tilde\nabla^e\tilde\nabla^f a_{|cf|}{}^d\tilde Y_{b)d}=-\tilde\Box \tilde Y_{ab}+\tilde\nabla^c\tilde\nabla_{(a}\tilde Y_{b)c}+
c_{(a}{}^{cde}\tilde\nabla^c\tilde\nabla^d\tilde Y_{b)}{}^e.
\eeq
Inserting this result into the above expression for $(DD\tilde Y)_{ab}$ we get
\beq\label{square27}
(DD\tilde Y)_{ab}=4(-\tilde\Box \tilde Y_{ab}+\tilde\nabla^c\tilde\nabla_{(a}\tilde Y_{b)c}-
c_{cde(a}\tilde\nabla^c\tilde\nabla^d\tilde Y_{b)}{}^e).
\eeq
Thus we have managed to eliminate the symmetric combination of covariant derivatives that has caused a bit of headache until now. The terms in the last
formula for $(DD\tilde Y)_{ab}$ can be dealt with rather easily as we now show. The first term is simply the coset master  equation squared, i.e., 
\beq
\tilde\Box \tilde Y_{ab}=-(C_G-C_H)\tilde Y_{ab}.
\eeq
The other two terms both involve the commutator of two covariant derivatives (since $\tilde\nabla^a\tilde Y_{ab}=0$). So we need to compute
\beq
[\tilde\nabla_c,\tilde\nabla_d]\tilde Y_{ab}=\tilde\nabla_c(\nabla_d\tilde Y_{ab}-\frac{m}{3}a_{da}{}^e\tilde Y_{eb}-\frac{m}{3}a_{db}{}^e\tilde Y_{ae})-(c \leftrightarrow d)=\nonumber
\eeq
\beq
\nabla_c\nabla_d\tilde Y_{ab}-\frac{m}{3}a_{cd}{}^e\nabla_e\tilde Y_{ab}-\frac{m}{3}a_{ca}{}^e\nabla_d\tilde Y_{eb}-\frac{m}{3}a_{cb}{}^e\nabla_d\tilde Y_{ae}
-\frac{m}{3}a_{da}{}^e\tilde\nabla_c\tilde Y_{eb}-\frac{m}{3}a_{db}{}^e\tilde\nabla_c\tilde Y_{ae}-(c \leftrightarrow d)=\nonumber
\eeq
\beq
[\nabla_c,\nabla_d]\tilde Y_{ab}-\frac{2m}{3}a_{cd}{}^e\nabla_e\tilde Y_{ab}
-\frac{2m^2}{9}(a_{[c|a|}{}^ea_{d]e}{}^f\tilde Y_{fb}+a_{[c|a|}{}^ea_{d]b}{}^f\tilde Y_{ef}+a_{[c|b|}{}^ea_{d]a}{}^f\tilde Y_{fe}+a_{[c|b|}{}^ea_{d]e}{}^f\tilde Y_{af}).
\eeq
Thus, all single derivative terms cancel except  one. Noting also that the two non-derivative terms in the middle of the bracket cancel the commutator becomes
\beq
[\tilde\nabla_c,\tilde\nabla_d]\tilde Y_{ab}=[\nabla_c,\nabla_d]\tilde Y_{ab}-\frac{2m}{3}a_{cd}{}^e\nabla_e\tilde Y_{ab}
-\frac{4m^2}{9}\,a_{[c}{}^{(a|e|}a_{d]e}{}^{|f|}\tilde Y_{f}{}^{b)}.
\eeq
Simplifying the non-derivative terms finally gives
\beq
[\tilde\nabla_c,\tilde\nabla_d]\tilde Y_{ab}=[\nabla_c,\nabla_d]\tilde Y_{ab}-\frac{2m}{3}a_{cd}{}^e\nabla_e\tilde Y_{ab}
+\frac{4m^2}{9}(\de_{(a}^{[c}\tilde Y_{b)}{}^{d]}-c_{cd(a}{}^e\tilde Y_{b)e}).
\eeq
Since we are interested in expressing the right hand side in terms of the $D$ operator defined above using $\tilde\nabla_a$ we 
rewrite the last equation as
\beq
[\tilde\nabla_c,\tilde\nabla_d]\tilde Y_{ab}=[\nabla_c,\nabla_d]\tilde Y_{ab}-\frac{2m}{3}a_{cd}{}^e\tilde\nabla_e\tilde Y_{ab}
-\frac{4m^2}{9}(\de_{(a}^{[c}\tilde Y_{b)}{}^{d]}+2c_{cd(a}{}^e\tilde Y_{b)e}).
\eeq
To get the final expression that will be useful here we replace the commutator on the right hand side by the Riemann tensor. This gives
\beq
[\tilde\nabla_c,\tilde\nabla_d]\tilde Y_{ab}=2W_{cd(a}{}^e\tilde Y_{b)e}-\frac{2m}{3}a_{cd}{}^e\tilde\nabla_e\tilde Y_{ab}+\frac{32m^2}{9}(\de_{(a}^{[c}\tilde Y_{b)}{}^{d]}-\frac{1}{4}c_{cd(a}{}^e\tilde Y_{b)e}).
\eeq

The first term we need to compute in the $DD\tilde Y$ equation contains the contracted expression $[\tilde\nabla^b,\tilde\nabla_d]\tilde Y_{ab}$. Setting $b=c$ in the last equation above  we get
\beq\label{symCommutatorContracted}
[\tilde\nabla^c,\tilde\nabla_a]\tilde Y_{bc}=W^c{}_{ab}{}^e\tilde Y_{ce}-\frac{2m}{3}a^c{}_{a}{}^e\tilde\nabla_e\tilde Y_{bc}+\frac{56m^2}{9}\tilde Y_{ab}.
\eeq
The second term in the $DD\tilde Y$ we need is the one with a  contraction of the commutator with the $c$ symbol:
\beq\label{cCommutatorContracted}
c_{cd(a}{}^f[\tilde\nabla_c,\tilde\nabla_d]\tilde Y_{b)f}=-2W_{(a}{}^e{}_{b)}{}^f\tilde Y_{ef}+\frac{8m}{3}a_{(a}{}^{ef}\tilde\nabla^e\tilde Y_{b)}{}^f+\frac{112m^2}{9}\tilde Y_{ab}.
\eeq

Inserting the  two results in \eqref{symCommutatorContracted} and \eqref{cCommutatorContracted} into \eqref{square27} we find
\beq
\frac{1}{4}(DD\tilde Y)_{ab}=-\tilde\Box \tilde Y_{ab}+\frac{m}{3}(D\tilde Y)_{ab}+\frac{112m^2}{9}\tilde Y_{ab}-2W_{(a}{}^e{}_{b)}{}^f\tilde Y_{ef}.
\eeq
Replacing the $G_2$ covariant box with Casimirs gives
\beq
\frac{1}{4}(DD\tilde Y)_{ab}=C_GY_{ab}+\frac{m}{3}(D\tilde Y)_{ab}+\frac{112m^2}{9}\tilde Y_{ab}-(C_H\tilde Y_{ab}+2W_{(a}{}^e{}_{b)}{}^f\tilde Y_{ef}).
\eeq
Then from the Appendix we know that the last bracket gives $\tfrac{112m^2}{9}$ which implies the amazingly simple equation
\beq\label{Squared3Form27}
(DD\tilde Y)_{ab}-\frac{4m}{3}(D\tilde Y)_{ab}-4C_G\tilde Y_{ab}=0.
\eeq
In view of the eigenvalue equation \eqref{27KappaEquation}, we may express the "solution" to the last equation as
\beq
D\tilde Y=\frac{2m}{3}\pm2\sqrt{C_G+\frac{m^2}{9}}=\frac{2m}{3}(1\pm\sqrt{20C_G+1}).
\eeq
Then by replacing  $(D\tilde Y)_{ab}$ with the expression coming from the 3-form \eqref{27KappaEquation}, we find
\beq
{\bf 27}:\,\,\,\,\kappa_3^2=\frac{m^2}{9}(20C_G+2\pm 2\sqrt{20C_G+1}).
\eeq
Since $\kappa_3^2$ is the eigenvalue of $\Delta_3=Q^2$ this  must be a square. Indeed, it can be written
\beq
{\bf 27}:\,\,\,\,\Delta_3=\frac{m^2}{9}(\sqrt{20C_G+1}\pm1)^2.
\eeq
\subsection{Lichnerowicz}

When we now turn to the transverse and traceless metric modes $h_{ab}$ on the squashed seven-sphere we can take advantage of 
the results obtained in the previous case of the 3-form. To see how to do this  let us write out the Lichnerowicz equation  explicitly
\beq
\Delta_L:\,\,\,\,\,\Delta_L h_{ab}=-\Box h_{ab}-2W_{acbd}h^{cd}+14m^2h_{ab}=\kappa_L^2h_{ab}.
\eeq
The coset master equation reads in this case
\beq
	 \nabla_a h_{bc}+\frac{2m}{3}a_{ad(b}h_{c)d}=-T_a h_{bc},
\eeq
which when squared gives
\beq
G/H:\,\,\,\,\,\,\Box h_{ab}+\frac{4m}{3}a_{cd(a}\nabla^c h_{b)}{}^d-\frac{14m^2}{9}h_{ab}=T_cT_ch_{ab}.
\eeq
Eliminating the box operator from the above $\Delta_L$ and $G/H$ equations and using $T_cT_c=-(C_H-C_H)$ gives
\beq
\kappa_L^2 h_{ab}=C_Hh_{ab}-(C_Hh_{ab}+2W_{acbd}h^{cd})+(14+\frac{14}{9})m^2h_{ab}+\frac{4m}{3}a_{cd(a}\tilde\nabla^c h_{b)}{}^d,
\eeq
where we have used the $G_2$ covariant derivative
\beq
\tilde\nabla_a h_{bc}=\nabla_a h_{bc}-\frac{2m}{3}a_{a(b}{}^dh_{c)}{}^d.
\eeq

As in the previous cases we now 
use the result for the 
Weyl tensor term from the Appendix, i.e., that $C_H+2W$ acting on the different $H$ irreps in ${\bf 27}$ gives the universal value $\tfrac{28}{5}=\tfrac{112}{5}m^2$. This gives us the rather simple equation
\beq\label{EigenvalueLic}
\kappa_L^2 h_{ab}=\tfrac{m^2}{9}(20C_G+28)h_{ab}+\frac{4m}{3}a_{cd(a}\tilde\nabla^c h_{b)}{}^d.
\eeq

There is one crucial difference between the Lichnerowicz modes and the 3-form modes analysed in the last subsection: Transversality of the metric modes does not imply that the associated 3-form $Y_{abc} \equiv \frac{4}{3}a_{d[ab}h_{c]d}$ is transverse (recall the result $\tilde \nabla^a\tilde Y_{ab}=0$ derived in the previous subsection). Thus when squaring $(Dh)_{ab}\equiv 2a_{cd(a}\tilde\nabla^c h_{b)}{}^d$ we cannot simply use \eqref{Squared3Form27} since this equation was derived with the assumption of a transverse $Y^{(27)}_{abc}$. The change is however relatively small; we only need to add the term in \eqref{SymmetricDerivative27} proportional to $\nabla^a Y^{(27)}_{abc}$ to proceed. We find
\beq\label{LicSquared}
	(DDh)_{ab}-\frac{4m}{3}(Dh)_{ab}-4C_G h_{ab}-8a_{cd(a}\tilde{\nabla}^c\tilde{\nabla}^eY^{(27)}_{|ed|b)} = 0.
\eeq
To deal with the last term we apply the $D$ operator yet another time.  To simplify the computation we introduce the notation $Y_{ab}\equiv \tilde{\nabla}^e Y_{eab} = \nabla^e Y_{eab}$ where $\nabla^aY_{ab}=0$ and define $(DY)_{ab} \equiv 2a_{cd(a}\tilde{\nabla}^cY_{|d|b)} = 2a_{cd(a}\nabla^cY_{|d|b)}$. It is then immediately found that
\beq
	(DDY)_{ab} = 2a_{cd(a}\tilde{\nabla}^{c}(DY)_{|d|b)} = -\frac{14m}{3}(DY)_{ab}+2a_{cd(a}\nabla^{c}(DY)_{|d|b)}
\eeq 
Expanding out the nontrivial last term on the right side gives
\beq\label{SquaredDivergence}
	2a_{cd(a}\nabla^c(DY)_{|d|b)} = 2a_{cd(a}\nabla^{c}a_{|efd}\nabla^{e}Y_{f|b)} + 2a_{cd(a}\nabla^{c}a_{|ef|b)}\nabla^{e}Y_{fd}\,.
\eeq

We start by analysing the first term in \eqref{SquaredDivergence}. When the first covariant derivative passes $a_{efd}$ we need to compute
\begin{equation}
\begin{aligned}
	a_{cd}{}^{(a}a_{efd}\nabla_c\nabla_e Y_{f}{}^{b)}&= -2\delta^{c(a}_{ef}\nabla_{c}\nabla_e Y_{f}{}^{b)}-c_{cef}{}^{(a}\nabla_c\nabla_eY_{f}{}^{b)} \\
	&= \nabla_{f}\nabla^{(a}Y_{f}{}^{b)} = 6mY^{(ab)}+R^{f(ab)g}Y_{fg}\\
	&=0
\end{aligned}
\end{equation}
In the second equality we have used the antisymmetry of $Y_{ab}$ and the fact that the 2-form projector $P_7$ vanishes when acting on the Weyl tensor (which implies that $c_{ab}{}^{cd}\propto \de_{ab}^{cd}$). 
The last equality is true due to the antisymmetry of $Y_{ab}$ together with the symmetrisation $(ab)$. When  the covariant derivative acts on  $a_{efg}$ we find
\beq
		ma_{cd}{}^{(a}c_{cefd}\nabla^{e}Y_{f}{}^{b)} = 4ma_{ef}{}^{(a}\nabla^{|e|}Y_{f}{}^{b)} = 2m(DY)_{ab}.
\eeq
We conclude that 
\beq
2a_{cd(a}\nabla^{c}a_{|efd}\nabla^{e}Y_{f|b)} = 4m(DY)_{ab}.
\eeq

We then turn to the second term in \eqref{SquaredDivergence}. Here the covariant derivative can act in two different ways. First, moving the derivative past $a_{efb}$ gives the expression
\beq
	a_{cd}{}^{(a}a_{ef}{}^{b)}\nabla^c \nabla^eY_{fd} = \thalf a_{cd}{}^{(a}a_{ef}{}^{b)}[\nabla^c, \nabla^e]Y_{fd}\,.
\eeq
The action of the antisymmetrized covariant derivatives will give two terms which are essentially identical and we only show how to deal with the first one. Using the squashed sphere Riemann tensor it follows that
\begin{equation}
\begin{aligned}
	a_{cd}{}^{(a}a_{ef}{}^{b)}R^{ce}{}_{fg}Y^g{}_d &= 2m^2\delta^{ce}_{fg}\,a_{cd}{}^{(a}a_{ef}{}^{b)}Y^{g}{}_d + a_{cd}{}^{(a}a_{ef}{}^{b)}W^{ce}{}_{fg}Y^{g}{}_d \\
	&= m^2 a_{fd}{}^{(a}a_{gf}{}^{b)}Y^g{}_{d}\\
	&=0,
\end{aligned}
\end{equation}
where we have used that $a_{efb}W^{ce}{}_{fg} = -\frac{1}{2}a_{efb}W^{cg}{}_{ef}=0$ (see the Appendix). The other term vanishes in exactly the same way. Finally, we have a term 
 coming from the covariant derivative hitting $a_{efb}$. After some algebra, this term becomes
\beq
	2ma_{cd}{}^{(a}c_{cef}{}^{b)}\nabla^e Y_{fd} = 2ma_{cd}{}^{(a}\nabla_c Y_{d}{}^{b)} = 2m(DY)_{ab},
\eeq
and so $a_{cd}{}^{(a}a_{ef}{}^{b)}\nabla^c \nabla^eY_{fd} = 2m(DY)_{ab}$. Putting these results together we find that
\beq
	(DDY)_{ab} = (6m-\frac{14m}{3})(DY)_{ab} = \frac{4m}{3}(DY)_{ab}.
\eeq

Having calculated the action of $D$ on $(DY)_{ab}$ we can apply $D$ on \eqref{LicSquared} and then subtracting the previous equation to eliminate the $(DY)_{ab}$ terms. From this procedure we find the third order equation
\beq
	(D-\frac{4m}{3})(D^2h-\frac{4m}{3}Dh-4C_Gh) = 0.
\eeq
"Solving" for $(Dh)_{ab}$ and plugging the result back into \eqref{EigenvalueLic} gives  the following three different eigenvalues:
\beq
	\Delta_L = \frac{m^2}{9}(20 C_G +36),\,\,\,\, \Delta_L = \frac{m^2}{9}\left(20C_G+32 \pm 4\sqrt{20C_G+1}\right).
\eeq

\section{Conclusions}

Let us  summarise what we know so far about the spectrum of operators on the squashed seven-sphere including the new results for $\Delta_2$ and $\Delta_3$ obtained in this paper.
The previously known eigenvalues for $\Delta_0$, $\slashed D_{1/2}$ \cite{Nilsson:1983ru}, $\Delta_1$ \cite{Yamagishi:1983ri},  and $\Delta_L$ \cite{Duff:1986hr} are
\beqa
\Delta_0&=&\tfrac{m^2}{9}\,20C_G,\\
{\slashed D}_{1/2}&=&-\tfrac{m}{2}\pm \tfrac{m}{3}\sqrt{20C_G+81},\\
{\slashed D}_{1/2}&=&\tfrac{m}{6}\pm \tfrac{m}{3}\sqrt{20C_G+49},\\
\Delta_1&=&\tfrac{m^2}{9}\,(20C_G+14\pm 2\sqrt{20C_G+49})=\tfrac{m^2}{9}\,(\sqrt{20C_G+49}\pm1)^2-4m^2,\\
\Delta_L&=&\tfrac{m^2}{9}\,(20C_G+36),\\
\Delta_L&=&\tfrac{m^2}{9}\,(20C_G+32\pm 4\sqrt{20C_G+1})=\tfrac{m^2}{9}\,(\sqrt{20C_G+1}\pm2)^2+3m^2,
\eeqa
while the new ones obtained in this paper are
\beqa
\Delta_2&=&\tfrac{m^2}{9}\,(20C_G+72),\\
\Delta_2&=&\tfrac{m^2}{9}\,(20C_G+44\pm 4\sqrt{20C_G+49})=\tfrac{m^2}{9}\,(\sqrt{20C_G+49}\pm2)^2-m^2,\\
\Delta_2&=&\tfrac{m^2}{9}\,20C_G,\\
\Delta_3&=&\tfrac{m^2}{9}\,(\sqrt{20C_G+49}\pm 1)^2,\\
\Delta_3&=&\tfrac{m^2}{9}\,(\sqrt{20C_G+81}\pm 3)^2,\\
\Delta_3&=&\tfrac{m^2}{9}\,(\sqrt{20C_G+1}\pm 1)^2.
\eeqa
Here it is appropriate to make some comments on the limitations of the obtained results. First, we have so far no results for the eigenvalues of 
the spin  $3/2$ operator ${\slashed D}_{3/2}$, although some can easily be extracted from supersymmetry. Secondly, for $\Delta_2$
and $\Delta_L$ we seem to lack some eigenvalues. This is indicated by the degeneracies in the cross diagrams for these two operators derived in \cite{Nilsson:2018lof},
as well as by supersymmetry. For $\Delta_3$ there may already be too many available eigenvalues but if one is supposed to pick only one sign when removing the
square (as done for ${\slashed D}_{1/2}$) one is instead lacking two eigenvalues. These problematic features are partly due to the fact that although we have extensive knowledge
of the eigenvalue spectra from the  list above, the method applied here does not provide  direct information how to associate these eigenvalues with the cross diagrams of  \cite{Nilsson:2018lof}.  
These and other issues will be elaborated upon in a forthcoming publication \cite{Nilsson:2021kn}. 

The results of this paper clearly demonstrate the important role of weak $G_2$ holonomy when solving the eigenvalue equations. There are, however, deeper issues in the  context of holonomy 
and string/M theory that might be interesting to study in relation to the squashed seven-sphere,  for instance the notion of generalised holonomy discussed in \cite{Batrachenko:2003ng}.

\section*{Acknowledgement}
One of us (BEWN) thanks M.J. Duff and C.N. Pope for a number of discussions on issues related to 
the theory analysed in this paper and for collaboration at an early stage of this project. 
We are also grateful to Joel Karlsson for some useful comments on the manuscript.
The work of S.E. is partially supported by the Knut and Alice Wallenberg Foundation under the grant: "Exact Results 
in Gauge and String Theories", Dnr KAW 2015.0083.

\appendix

\section{Octonions:  Conventions and some identities}

The octonions satisfy a non-associative algebra defined by the   totally antisymmetric structure constants $a_{abc}$ ($a,b,c,..=0,1,...,6$):
\beq
a_{abc}=1\,\,\text{for}\,\,abc=456, 041, 052, 063, 162, 135, 243.
\eeq
By splitting $a=(0,i,\hat{i})$ (with $i=1,2,3,\hat{i}=4,5,6=\hat 1, \hat 2, \hat 3$) they can be written more compactly as 
\beq
	a_{0i\hat{j}} = -\delta_{ij},
	\quad
	a_{ij\hat{k}} = -\epsilon_{ijk},
	\quad
	a_{\hat{i}\hat{j}\hat{k}} = \epsilon_{ijk}.
\eeq
We define its dual, denoted $c_{abcd}$, by
\beq
c_{abcd}=\tfrac{1}{6}\ep_{abcdefg}\,a_{efg},
\eeq
where the epsilon tensor is totally antisymmetric with $\ep_{0123456}=1$.

Sometimes  it is  convenient  to write the  gamma matrices $\Gamma_a$ in seven dimensions
in terms of the structure constants defined above and a Killing spinor  $\eta$ satisfying 
\beq
\nabla_a\,\eta=-\frac{i}{2}m\Gamma_a\,\eta,\,\,\,\bar\eta\,\eta=1.
\eeq
The conventions used here are 
\beq
a_{abc}=i\bar\eta\Gamma_{abc}\eta,\,\,\,\,c_{abcd}=-\bar\eta\Gamma_{abcd}\eta,
\eeq
and 
\beq
\{\Gamma_a,\Gamma_b\}=2\de_{ab},\,\,\,-i\Gamma_{abcdefg}=\ep_{abcdefg}\bf{1}.
\eeq
These gamma matrices  are $8\times 8$ matrices with spinor indices $A,B,...$ taking the eight values $A=(a,8)$ etc.
The consistency of these conventions can then be verified by explicitly writing out the gamma matrices  in terms of 
the structure constants as follows\footnote{Note that the indices for both vectors,  $a,b,c,...$,  and spinors,  $A, B, C, ...$,  are raised and lowered by a unit matrix which means that equations may appear  with indices 
in the wrong up or down position.}:
\beq
(\Gamma_a)_B{}^C: \\
(\Gamma_a)_b{}^c=i a_{abc},\,\,\,\,
(\Gamma_a)_b{}^8=i \de_{ab}, \,\,\,\,
(\Gamma_a)_8{}^b=-i \de_{ab}.
\eeq
\beq
(\Gamma_{ab})_B{}^C: \\
(\Gamma_{ab})^{cd}= 2\delta^{cd}_{ab}-c_{ab}{}^{cd},\,\,\,\,
(\Gamma_{ab})_c{}^8= a_{abc}.
\eeq
Note that all seven gamma matrices are antisymmetric and imaginary. With these definitions one can check that $\bar\eta\,\Gamma_{abc}\,\eta=-ia_{abc}$ as stipulated above. However, this calculation requires some octonionic structure constant identities that are seldom given in the literature. For the convenience of the reader we list the identities used in this paper below:
\beq
a_{abe}\,a^{cde}=2\de_{ab}^{cd}+c_{ab}{}^{cd},
\eeq
\beq
c_{abc}{}^{g}c^{defg}+a_{abc}a^{def}=6\de_{abc}^{def}+9\de_{[a}^{[d}c_{bc]}{}^{ef]},
\eeq
\beq
c_{abcf}a^{def}=6a_{[ab}{}^{[d}\de^{e]}_{c]}.
\eeq
These identities may be verified directly by inserting the actual values of the constants, or by using the above expressions in terms of the Killing spinor $\eta$ and the gamma matrices. The latter approach requires the Fierz formula
\beq
\Gamma_a\eta\bar\eta\Gamma_a={\bf 1}-\eta\bar\eta.
\eeq
Various contractions of indices in these identities lead directly to a number of  additional identities that are used frequently in the main text.  A quite extensive list of identities is presented in \cite{House:2004pm}.

Using the Killing spinor equation we furthermore find the derivative identities
\beq
	\nabla_{a}a_{bcd} = -\frac{m}{2}\bar\eta(\Gamma_a \Gamma_{bcd}-\Gamma_{bcd}\Gamma_{a})\eta = -m\bar\eta(\Gamma_{abcd})\eta = mc_{abcd},
\eeq
\beq
	\nabla_{a}c_{bcde} = -\frac{i m}{2}\bar \eta(\Gamma_{a}\Gamma_{bcde}-\Gamma_{bcde}\Gamma_{a})\eta = -4im\delta_{a[b}\bar \eta(\Gamma_{cde]})\eta = -4m\delta_{a[b}a_{cde]}.
\eeq

\section{Projection operators}
\subsection{2-forms }
\beq
(P_7)_{ab}{}^{cd}=\frac{1}{6}a_{ab}{}^ea^{cde}=\tfrac{1}{6}(2\de_{ab}^{cd}+c_{ab}{}^{cd}),
\eeq
\beq
(P_{14})_{ab}{}^{cd}=\frac{1}{6}(4\de_{ab}^{cd}-c_{ab}{}^{cd}),
\eeq
which implies the useful relation
\beq
c_{ab}{}^{cd}=2(2P_7-P_{14})_{ab}{}^{cd}.
\eeq
\subsection{3-forms}
\beq
(P_1)_{abc}{}^{def}=\frac{1}{42}a_{abc}a^{def},
\eeq
\beq
(P_{7})_{abc}{}^{def}=\frac{1}{24}c_{abc}{}^{g}c^{defg}=\frac{1}{24}(6\de_{abc}^{def}+9\de_{[a}^{[d}c_{bc]}{}^{ef]}-a_{abc}a^{def}),
\eeq
\beq
(P_{27})_{abc}{}^{def}=(1-P_1-P_7)_{abc}{}^{def}=\frac{1}{56}(42\de_{abc}^{def}-21\de_{[a}^{[d}c_{bc]}{}^{ef]}+a_{abc}a^{def}).
\eeq

\subsection{Spin $1/2$ }

The purpose of the projection operators in this case is to split the $SO(7)$ spinor representation into two $G_2$ irreps as follows
${\bf 8}\rightarrow {\bf 1} \oplus {\bf 7}$:
\beq
P^s_1=\frac{1}{8}({\bf 1}-\frac{1}{24}c_{abcd}\Gamma^{abcd}),
\eeq
\beq
P^s_7={\bf 1}-P^s_1=\frac{1}{8}(7\cdot {\bf 1}+\frac{1}{24}c_{abcd}\Gamma^{abcd}).
\eeq
While they trivially sum to the unit matrix, the fact that they both square to themselves and are orthogonal to each other require some algebra to show.
In fact, all these properties follow directly if we show that $(P^s_1)^2=P_1^s$. This is done as follows
\beq
(P^s_1)^2=\frac{1}{64}({\bf 1}-\frac{1}{12}c_{abcd}\Gamma^{abcd}+\frac{1}{(24)^2}c_{abcd}c_{efgh}\Gamma^{abcd}\Gamma^{efgh})=P_1^s.
\eeq
The  key calculation in the last step is to expand $\Gamma^{abcd}\Gamma^{efgh}$ in the gamma basis $\Gamma^{(n)}$, $n=0,....,7$, in seven dimensions:
\beq
\Gamma^{abcd}\Gamma_{efgh}=-16\de^{[a}_{[e}\Gamma^{bcd]}{}_{fgh]}-72\de^{[ab}_{[ef}\Gamma^{cd]}{}_{gh]}+96\de^{[abc}_{[efg}\Gamma^{d]}{}_{h]}+24\de^{abcd}_{efgh}.
\eeq
However, when contracted with $c_{abcd}c_{efgh}$ the first and third terms vanish since they are antisymmetric under interchange of the two index sets $abcd$ and $efgh$.
The remaining two terms are  easily computed using the two identities for $c_{abcd}$: $c^{abcd}c_{efcd}=8\de^{ab}_{ef}+2c^{ab}{}_{ef}$ and $c^{abcd}c_{abcd}=168$.
 $(P^s_1)^2=P_1^s$ then follows directly.

\section{The role of the Weyl tensor}

The Weyl tensor of the squashed seven-sphere is given in \cite{Duff:1986hr}. The explicit expression given there can be written 
in a more compact form in terms of 't Hooft symbols. By setting $\alpha = (0,1,2,3)=(0,i)$ and introducing the 't Hooft symbols by
\beqa
	\eta^k_{\alpha\beta}:&&\,\,\,\, \eta^k_{ij}=\epsilon_{ijk},\,\,\,\,\,\eta^k_{0j}=-\eta^k_{j0}=-\de_{jk},\\
	\bar{\eta}^i_{\alpha\beta}:&& \,\,\,\, \bar\eta^k_{ij}=\epsilon_{ijk},\,\,\,\,\,\bar\eta^k_{0j}=-\bar\eta^k_{j0}=\de_{jk},
\eeqa
 the Weyl tensor can be  explicitly expressed as follows
\beq
W_{\al\be}{}^{\ga\de}=\frac{4}{5}\de_{\al\be}^{\ga\de},\,\,\,W_{\hat i \hat j}{}^{\hat k \hat l}=\frac{8}{5}\de_{i j}^{kl},\,\,\,\,
W_{\al\be}{}^{\hat k \hat l}=\frac{2}{5}\bar\eta_{\al\be}^m\ep^{mkl},\,\,\,\,W_{\al}{}^{\hat j}{}_{\ga}{}^{\hat l}=\frac{1}{5}\bar\eta_{\al\ga}^m\ep^{mjl}-\frac{2}{5}\de_{\al\ga}\de^{ j l}.
\eeq
The demonstration that the squashed seven-sphere has holonomy $G_2$  relies on the fact that defining $W_{ab}\equiv \tfrac{1}{4}W_{ab}{}^{cd}\Gamma_{cd}$ we find
\beq
W_{0i}=\tfrac{1}{5}(\Gamma_{0i}+\thalf\ep_{ijk}\Gamma_{\hat j \hat k}),
\eeq
\beq
W_{ij}=\tfrac{1}{5}(\Gamma_{ij}+\Gamma_{\hat i \hat j}),
\eeq
\beq
W_{i\hat j}=\tfrac{1}{5}(-\Gamma_{i\hat j}-\thalf\Gamma_{j \hat i}+\thalf\de_{ij}\Gamma_{k \hat k}-\thalf\ep_{ijk}\Gamma_{0 \hat k}),
\eeq
\beq
W_{0\hat i}=-\tfrac{1}{5}(\Gamma_{0\hat i}+\thalf\epsilon_{ijk}\Gamma_{j \hat k}),
\eeq
\beq
W_{\hat i \hat j}=\tfrac{1}{5}(2\Gamma_{\hat i \hat j}+\Gamma_{i j}+\epsilon_{ijk}\Gamma_{0k}),
\eeq
while the remaining components are not independent but instead given by
\beq
W_{i\hat i}=0,\,\,\,\,\,W_{0\hat i}=\ep_{ijk}W_{j\hat k},\,\,\,W_{\hat i \hat j}=W_{ij}+\ep_{ijk}W_{0k}.
\eeq
Thus only 14 components are linearly independent which was shown in \cite{Duff:1986hr} to result in $G_2$ holonomy. An immediate consequence, as can be easily verified using the explicit expressions, is that
\beq
	a_{abc}W_{bc}{}^{de} = 0\,.
\eeq 
To obtain identities of this kind it is convenient to express the octonionic structure constants in terms of 't Hooft symbols. The non-zero components are then (recall that $a=(\al, \hat i)$ and $\hat i=(\hat 1, \hat 2, \hat 3)=(4,5,6)$):
\beqa
a_{abc}&&:\,\,\,\,a_{\al\be\hat k}=-\bar\eta_{\al\be}^{k},\,\,\,\,\, a_{\hat i \hat j \hat k}=\ep_{ijk} , \\
c_{abcd}:&&\,\,\,\,c_{\al\be\ga\de}=\ep_{\al\be\ga\de}=\bar\eta_{[\al\be}^{k}\bar\eta_{\ga\de]}^{k},\,\,\,\,\, c_{\al\be\hat i \hat j}=-\bar{\eta}^k_{\alpha\beta}\ep_{ijk},
\eeqa
where $\ep_{\al\be\ga\de}$ is antisymmetric with $\ep_{0123}=1$.

In our present context we will for instance
use these results to express the 2-form modes $Y^{(14)}_{ab}$ in a way that makes its projections onto the $H$ irreps clear, i.e., the decomposition
\beq
G_2\rightarrow Sp_1^A\times Sp_1^{B+C}:\,\,\,\,\,\, {\bf 14}\rightarrow (1,3)\oplus (2,0) \oplus (0,2).
\eeq
To this end we will parametrise the $Spin(4)\times SU(2)$ subgroup 
of $G=Sp_2\times Sp_1^C$ using
\beq
\Gamma^a=(\Gamma^{\al},\Gamma^{\hat i}):\,\,\,\,\,\,\,\,\,\,\,\,
\Gamma^{\al}=
\begin{pmatrix}
0 & (\si^{\al})_{A\dot B} \\
(\bar \si^{\al})^{\dot A B} & 0  \\
\end{pmatrix}
\otimes \de^{\dot A}_{\dot B},\,\,\,\,\,\Gamma^{\hat i}={\bf 1}_{4\times 4}\otimes (\si^{\hat i})^{\dot A}{}_{\dot B},
\eeq
where
\beq
\si^{\al}\equiv (-i{\bf 1}, \si^i),\,\,\,\bar\si^{\al}\equiv (i{\bf 1}, \si^i).
\eeq
Note the dot notation on the indices in the second tensor factor. This reflects the fact that we are using the  diagonal subgroup 
of $Sp_1^B$ and $Sp_1^C$ where the  former is a subgroup of $Sp_2$ when split into $Sp_1^A\times Sp_1^B$ according to the first
tensor factor.

The 't Hooft symbols $(\eta_{\al\be}^m , \bar \eta_{\al\be}^m)$ introduced above  can also be defined from the Pauli matrices by
\beq\label{etatosigma}
\si^{\al\be}\equiv i\eta_{\al\be}^m\si^m,\,\,\,\,\,
\bar\si^{\al\be}\equiv i\bar\eta_{\al\be}^m\si^m.
\eeq

\subsection{2-form modes }

From the above expressions we immediately obtain the projectors onto irreps $(s,t)$ of $H=Sp_1^A\times Sp_1^{B+C}$
\beq
(P^{ab}_{{\bf 14}\rightarrow (2,0)}Y_{ab})_A{}^B=(\si^{\al\be}Y_{\al\be})_A{}^B,
\eeq
\beq
(P^{ab}_{{\bf 14}\rightarrow (0,2)}Y_{ab})^{\dot A}{}_{\dot B}=(\bar\si^{\al\be}Y_{\al\be}+2\si^{\hat i\hat j}Y_{\hat i\hat j})^{\dot A}{}_{\dot B},
\eeq
\beq
(P^{ab}_{{\bf 14}\rightarrow (1,3)}Y_{ab})_{A(\dot B\dot C \dot D)}=(\si^{\al})_{A(\dot B}\si^{\hat j}_{\dot C \dot D)}Y_{\al \hat j}.
\eeq
As  two examples of how to use these projectors we compute the Weyl tensor eigenvalue of the $(2,0)$ and $(0,2)$ irreps. For $(2,0)$
\beq
P_{{\bf 14}\rightarrow (2,0)}^{ab}W_{abcd}Y_{cd}=\si^{\al\be}W_{\al\be cd}Y_{cd}=\si^{\al\be}(\tfrac{4}{5}\de_{\al\be}^{\ga\de}Y_{\ga\de}+\tfrac{2}{5}\bar\eta^m_{\al\be}\ep^{mkl}Y_{\hat k\hat l})=\tfrac{4}{5}\si^{\al\be}Y_{\al\be},
\eeq
showing that the sought for eigenvalue is $\tfrac{4}{5}$. Note that in the last equality we used the  fact that $\si^{\al\be}\bar\eta^m_{\al\be}=0$. For $(0,2)$ we find
\beq\label{eq:021}
	2\si^{\hat i \hat j} W_{\hat i \hat j c d}Y^{cd} = 2\si^{\hat i\hat j}(\frac{8}{5}Y_{\hat i \hat j} + \frac{2}{5}\bar{\eta}^m_{\alpha\beta}\epsilon^{m\hat i \hat j}Y^{\alpha\beta}) = \frac{8}{5}(2\si^{\hat i \hat j}Y_{\hat i \hat j}+\bar{\si}^{\alpha\beta}Y_{\alpha\beta})
\eeq
\beq\label{eq:022}
\bar{\si}^{\alpha\beta}W_{\alpha\beta a b}Y^{ab} = \bar{\si}^{\alpha\beta}(\frac{4}{5}Y_{\alpha\beta}+\frac{2}{5}\bar{\eta}^{m}_{\alpha\beta}\epsilon^{mij}Y_{\hat i \hat j}) = \frac{4}{5}(\bar{\si}^{\alpha\beta}Y_{\alpha\beta}+2\si^{\hat i \hat j}Y_{\hat{i}\hat{j}})
\eeq
where we have used \eqref{etatosigma}. Then adding together \eqref{eq:021} and \eqref{eq:022} implies $P_{\mathbf{14}\rightarrow (0,2)}WY = \frac{12}{5}P_{\mathbf{14}\rightarrow (0,2)}Y$. Repeating this for the last case one finds the values tabulated below together with
the corresponding values of the $H$ Casimir operator $C_H$.

\begin{table}[H]
    \centering
    \begin{tabular}{|c|c|c|c|}
    \hline
         H-irrep & H Casimir & Weyl eigenvalue & Sum  \\
         \hline
         \hline
    $(2,0)$ & $4$ & \ \ $\frac{4}{5}$ \ \ & $\frac{24}{5}$ \\
    $(0,2)$ & $\frac{12}{5}$ & \ \ $\frac{12}{5}$ \ \ & $\frac{24}{5}$ \\
    $(1,3)$ & $6$ & $-\frac{6}{5}$ \ \ & $\frac{24}{5}$ \\
    \hline  
    \end{tabular}
    \caption{Weyl tensor eigenvalues for the $H$ irreps of the $G_2$ ${\bf 14}$-part of the 2-form.}
    \label{table:weyleigenvaluestwoforms}
\end{table}

Note that  the two eigenvalues do not individually respect the $G_2$ holonomy but that their sums do. Similarly, we know from before that
the Weyl tensor does not enter the equations for the $G_2$ ${\bf 7}$-part of the 2-form but that the $H$ irreps both give $\tfrac{12}{5}$ which thus automatically respects $G_2$.
\subsection{3-form and metric modes: the {\bf 27} irrep}

Here we need two more projectors, namely\footnote{In the first formula below the use of the symmetrisation brackets is not optimal: The two pairs of dotted and undotted indices are symmetrised independently
of each other.}
\beq
(P^{ab}_{{\bf 27}\rightarrow(2,2)}Y_{ab})_{(A(\dot B C)\dot D)}=(\bar \si^{\al})_{(A(\dot B}(\bar \si^{\be})_{C)\dot D)}Y_{\al\be},
\eeq
\beq
(P^{ab}_{{\bf 27}\rightarrow(0,4)}Y_{ab})_{(\dot A\dot B\dot C\dot D)}=(\bar \si^{\hat i})_{(\dot A\dot B}(\bar \si^{\hat j})_{\dot C\dot D)}Y_{\hat i \hat j}.
\eeq
As an example how to get the Weyl eigenvalues we consider the second projector. The calculation works as follows:
\beq
(P^{ab}_{{\bf 27}\rightarrow(0,4)}W_{acbd}h^{cd})=\si^{\hat i}\si^{\hat j}W_{\hat i c \hat j d}h^{cd}=\si^{\hat i}\si^{\hat j}(\tfrac{8}{5}\de^{\hat i \hat k}_{\hat j \hat l}h^{\hat k \hat l}+\tfrac{1}{5}\bar \eta^m_{\al\be}\ep_{m\hat i \hat j}h^{\al\be}
-\tfrac{2}{5}\de_{\hat i \hat j}h_{\al\al}).
\eeq
The term containing the 't Hooft symbol vanishes since $\bar \eta^m_{\al\be}$ is antisymmetric in the two lower indices. The remaining terms simplify immediately to 
\beq
(P^{ab}_{{\bf 27}\rightarrow(0,4)}W_{acbd}h^{cd})=-\tfrac{4}{5}(\si^{\hat i}\si^{\hat j})_{(0,4)}h_{\hat i \hat j},
\eeq
where we also used that tracing over the hatted indices on $\si^{\hat i}\si^{\hat j}$ gives zero since this expression is in the irrep ${\bf 5}$ of $SU(2)$ given by the indices $(\dot A \dot B \dot C \dot D)$.

\begin{table}[H]
    \centering
    \begin{tabular}{|c|c|c|c|}
    \hline
         H-irrep & H Casimir & 2 $\times$ Weyl eigenvalue & Sum  \\
         \hline
         \hline
    $(0,0)$ & $0$ & \ \ $\frac{28}{5}$ \ \ & $\frac{28}{5}$ \\
    $(2,2)$ & $\frac{32}{5}$ & $-\frac{4}{5}$ \ \ & $\frac{28}{5}$ \\
    $(0,4)$ & $\frac{36}{5}$ & $-\frac{8}{5}$ \ \ & $\frac{28}{5}$ \\
    $(1,1)$ & $\frac{12}{5}$ & \ \ $\frac{16}{5}$ \ \ & $\frac{28}{5}$ \\
	$(1,3)$ & $6$ & $-\frac{2}{5}$ \ \ & $\frac{28}{5}$ \\
    \hline  
    \end{tabular}
    \caption{Weyl tensor eigenvalues for the $H$ irreps of the $G_2$ ${\bf 27}$-part of the 3-form.}
    \label{table:weyleigenvaluesLic}
\end{table}

Again we find the striking result that all $H$ irrep pieces a  $G_2$ irrep produce the same value for $C_H+2\times Weyl$, this time for the ${\bf 27}$.

In the main text we found the expression $C_HY_{abc} - 6W^d{}_{[ab}{}^{e}Y_{c]de}$. Using the explicit form of the projection operators we can verify that $W^{d}{}_{[ab}{}^{e}Y_{c]de}=W^{d}{}_{[ab}{}^{e}Y^{(27)}_{c]de}$. Applying $(\tilde P^{27})_{ab}{}^{fgh} \equiv a_{acd}(P^{27})_{cdb}{}^{fgh}$ gives
\beq
	(\tilde P^{27})_{ab}{}^{fgh}W^{d}{}_{fg}{}^{e}Y^{(27)}_{gde} = -\frac{1}{3}W_{(a}{}^{c}{}_{b)}{}^{d}\tilde{Y}_{cd}.
\eeq 
Using the eigenvalues listed in Table \ref{table:weyleigenvaluesLic} it follows that
\begin{equation}
	(\tilde P^{27})_{ab}{}^{fgh}(C_HY_{abc} - 6W^d{}_{[ab}{}^{e}Y_{c]de}) = (C_H\tilde Y_{ab} + 2W_{(a}{}^{c}{}_{b)}{}^e\tilde{Y}_{ce})= \frac{28}{5}\tilde{Y}_{ab}.
\end{equation}
Going back to the 3-form notation and using that $C_H$ has eigenvalue $\frac{12}{5}$ on $\mathbf{7}$ we find
\beq
	C_HY_{abc} - 6W^d{}_{[ab}{}^{e}Y_{c]de} = -\frac{1}{10}c_{abcd}Y_d +\frac{28}{5}Y_{abc}.
\eeq

\end{document}